\documentstyle[12pt]{article}
\newcommand{\be}{\begin{equation}}
\newcommand{\ee}{\end{equation}}
\newcommand{\ba}{\begin{eqnarray}}
\newcommand{\ea}{\end{eqnarray}}
\newcommand{\by}{\begin{eqnarray*}}
\newcommand{\ey}{\end{eqnarray*}}

\newcommand{\e}{\epsilon}
\newcommand{\ve}{\varepsilon}
\newcommand{\p}{\partial}
\newcommand{\ra}{\rightarrow}
\newcommand{\La}{\Lambda}

\newcommand{\Om}{\Omega}
\newcommand{\om}{\omega}
\newcommand{\al}{\alpha}

\newcommand{\del}{\delta}
\newcommand{\si}{\sigma}
\newcommand{\Si}{\Sigma}
\newcommand{\f}{\frac}
\newcommand{\ti}{\tilde}

\newcommand{\ct}{\cite}

\newcommand{\Ga}{\Gamma}
\newcommand{\scs}{\scriptstyle}

\newcommand{\ts}{\times}
\newcommand{\ul}{\underline}
\newcommand{\nn}{\nonumber}

\def\beq{\begin{equation}}
\def\eeq{\end{equation}}
\def\bea{\begin{eqnarray}}
\def\eea{\end{eqnarray}}
\def\bq{\begin{quote}}
\def\eq{\end{quote}}

\def\gappeq{\mathrel{\rlap {\raise.5ex\hbox{$>$}}
{\lower.5ex\hbox{$\sim$}}}}

\def\lappeq{\mathrel{\rlap{\raise.5ex\hbox{$<$}}
{\lower.5ex\hbox{$\sim$}}}}

\begin{document}
\pagestyle{empty}
\begin{flushright}
{CERN-TH/99-113}
\end{flushright}
\vspace*{5mm}
\begin{center}
{\bf MAGNETIC OSCILLATIONS IN DENSE COLD QUARK MATTER
 WITH FOUR-FERMION INTERACTIONS} \\
\vspace*{1cm} 
{\bf D. Ebert}$^\ast$\ ,
{\bf K.G. Klimenko}$^\ddagger$\ ,
{\bf M.A. Vdovichenko}$^\dagger$ and \fbox{{\bf A.S. Vshivtse}v$^\dagger$}

\vspace{0.2cm} 
$^{\ast}${\it Theory Division, CERN, CH 1211 Geneva 23, Switzerland\\
and Institut f\"ur Physik, Humboldt-Universit\"at, D-10115 Berlin, Germany
e-mail:~dietmar.ebert@cern.ch} \\

$^{\ddagger}${\it Institute for High Energy Physics, 142284 Protvino,
Moscow Region, Russia 
e-mail:~kklim@mx.ihep.su} \\

$^{\dagger}${\it Moscow Institute of Radio-Engineering, Electronics and 
Automatic Systems,\\
117454 Moscow, Russia} \\

\vspace*{2cm}  
{\bf ABSTRACT} \\ \end{center}
\vspace*{5mm}
\noindent
The phase structures of Nambu--Jona-Lasinio models with one or
two flavours have been investigated at non-zero values of
$\mu$ and $H$, where $H$ is an external magnetic field and $\mu$ is
the chemical potential. In the phase portraits of both models
there arise infinitely many massless chirally symmetric phases,
as well as massive ones with spontaneously broken chiral
invariance, reflecting the existence of infinitely many Landau
levels.  Phase transitions of first and second orders and a lot of
tricritical points have been shown to exist in phase diagrams.
In the massless case, such a phase structure leads unavoidably to
the standard van Alphen--de Haas magnetic oscillations of some
thermodynamical quantities, including magnetization, pressure and
particle density. In the massive case we have found an
oscillating behaviour not only for thermodynamical quantities,
but also for a dynamical quantity as the quark mass. Besides,
in this case we have non-standard, i.e. non-periodic, magnetic
oscillations, since the frequency of oscillations is an
$H$-dependent quantity.

\vspace*{2cm} 


\begin{flushleft} CERN-TH/99-113 \\
April 1999
\end{flushleft}
\vfill\eject

\setcounter{page}{1}
\pagestyle{plain}

\vspace*{0.6cm}

\section*{1. Introduction}

The exploration of strongly interacting matter at high density and in the 
presence of external electromagnetic fields is of fundamental
interest and has potential applications to the quark--gluon plasma and 
heavy-ion collisions, to cosmology and astrophysics of neutron stars.
Recently,  some aspects of this problem were considered in \cite{alford},
where it was pointed out that in QCD at high density a new phase
with  colour superconductivity might exist. The influence of external magnetic
fields on the QCD vacuum was, for example, studied in \ct{smilga}.

Our goal is to investigate properties of the strongly
interacting cold quark matter in the presence of
both external magnetic field $H$ and non-zero chemical potential
$\mu$.  The subject is closely related to magnetic oscillations
of different physical quantities. In this connection we should
remember that the van Alphen--de Haas effect (oscillations of
the magnetization) was for the first time predicted by Landau
and then experimentally observed in some non-relativistic systems
(in metals) more than sixty years ago \ct{haas,lan}.  At present,
a lot of the attention of researchers dealing with magnetic
oscillations is focused on relativistic condensed matter systems
(mainly on QED at $\mu,H\neq 0$), since the results of these
studies may be applied to cosmology, astrophysics and
high-energy physics \cite{per,as}.

It is well known that up to now the consideration of QCD at
$\mu,H\neq 0$ is a difficult problem. This is partly due to the
fact that numerical lattice simulations at $\mu\neq 0$ have not
been able to overcome problems associated with the complex part
of the fermionic determinant. Moreover, the incorporation of a
magnetic field into  lattice gauge calculations is not
elaborated sufficiently, either.  For these reasons, 
when considering quark matter at $\mu,H\neq 0$,
many authors prefer to deal with adequate models (e.g.
with the MIT bag model \ct{chakr}), rather than with QCD.

In the present paper we shall study
the above problem in the framework of some specific QCD-like quark
models. Namely, we shall investigate the influence of an
external magnetic field and chemical potential on the vacuum
structure of Nambu--Jona-Lasinio (NJL) models containing
four-fermion interactions \ct{nam,vaks}.  The simplest one,
denoted as model I, refers to the one-flavour case and is
presented by the Lagrangian

\be
\label{eq.1}
L_1=\bar{q}_k i \hat {\partial} q_k
+\frac{G}{2N_c}
[(\bar{q}_k q_k)^2
+(\bar{q}_k i\gamma_5 q_k)^2],
\ee
where all quark fields belong to the fundamental multiplet of
the colour $SU(N_c)$ group (here the summation over the colour index
$k=1,...,N_c$ is implied). Obviously, $L_1$ is invariant under
(global) $SU(N_c)$ and $U(1)_V$ transformations as well as
continuous $U(1)_A$ chiral transformations
\be
\label{eq.2}
q_k\to e^{i\theta\gamma_5}q_k~~;~~ (k=1,...,N_c).
\ee

The second, more realistic case considered here and referred to
as model II is a two-flavour NJL model whose Lagrangian has the
form
\begin{eqnarray}
\label{eq.3}
L_2&=& \bar q i\hat\p
q + \frac{G}{2N_c}[(\bar qq)^2+(\bar qi\gamma ^5\vec\tau q)^2],
\end{eqnarray}
where $q$ is a flavour isodoublet and colour-$N_c$-plet quark
field and  $\vec\tau$ are isospin Pauli matrices (in (3) and
below, flavour and colour indices of the quark field $q$ are now
suppressed).  The Lagrangian $L_2$ is invariant under (global)
$U(2)_f\ts SU(N_c)$ as well as under chiral $U(2)_L\ts U(2)_R$
groups.

NJL models were proposed as a good laboratory for investigating
the non-perturbative phenomenon of dynamical chiral symmetry
breaking (DCSB), which occurs in the physics of strong
interactions, as well as for describing the low-energy sector of
QCD (see e.g. papers \ct{ebert,volkov, kl} and references therein).
Since there are no closed physical systems in nature, the influence of
different external factors on the DCSB mechanism is of great interest.
In this relation, special attention has been paid to the analysis of the
vacuum structure of NJL-type models at non-zero temperature and chemical
potential \ct{kaw,1}, in the presence of external (chromo-)magnetic fields
\ct{klev,eb,gus}, with allowance for curvature and non-trivial space-time
topology \ct{In,2}. The combined influence of external electromagnetic
and gravitational fields on the DCSB effect in four-fermion field theories
was investigated in \ct{muta,od}.

In the present paper the phase structures and related
oscillating effects of the above-mentioned  Nambu--
Jona-Lasinio models are considered at $\mu,H\neq 0$. We will
show that, here, the set of oscillating physical parameters in NJL
models is richer than in QED at $\mu,H\neq 0$. Besides, in the
NJL models, in contrast to QED and similar to some condensed-
matter materials, there exist non-periodic magnetic oscillations.

\section*{2. NJL models at $\mu\neq 0$ and $H=0$}

First of all let us prepare the basis for the investigations in the following
sections and consider in detail the phase structure of the model
I at non-zero chemical potential $\mu\neq 0$ and $H=0$.

Recall some well-known vacuum properties
of the theory (1) at $\mu=0$. The introduction of an intermediate 
quark--meson Lagrangian
\be
\tilde L_1=\bar q i\hat\partial q-
\bar q (\sigma_1+i\sigma_2\gamma_5)
q-\frac{N_c}{2G}(\sigma^2_1+\sigma^2_2)
\ee
greatly facilitates the problem under consideration. 
(In (4) and other formulae below we have omitted 
the fermionic index $k$ for simplicity.) 
Clearly, using the equations of motion for the bosonic
fields $\sigma_{1,2}$, the theory in (4) is equivalent to that in (1).
From (4) we obtain the one-loop expression for the effective action
$$
\exp (iN_c S_{eff}(\sigma_{1,2})) =
\int D\bar q Dq \exp\left(i\int\tilde L_1 d^4x\right),
$$
where
\[
S_{eff}(\sigma_{1,2})=-
\int d^4x\frac{\sigma_1^2+\sigma^2_2}{2G}-i\ln\det
(i\hat\partial-\sigma_1-i\gamma_5\sigma_2).
\]
Assuming that in this formula $\sigma_{1,2}$ are independent of 
space-time points, we have by definition
\[
S_{eff}(\sigma_{1,2})=-
V_{0}(\sigma_{1,2})\int d^4 x,
\]
\be
V_{0}(\sigma_{1,2})=
\frac{\Sigma^2}{2G}+2i
\int \frac{d^4p}{(2\pi)^4}\ln
(\Sigma^2-p^2)\equiv V_0(\Sigma),
\ee
where $\Sigma=\sqrt{\sigma^2_1+\sigma^2_2}$.
Next, introducing in (5) Euclidean metric $(p_0\to ip_0)$ and cutting 
off the range
of integration $(p^2\leq \Lambda^2)$, we obtain
\be
V_0(\Sigma)=
\frac{\Sigma^2}{2G}-
\frac{1}{16\pi^2}\Biggl\{
\Lambda^4\ln
\left(1+\frac{\Sigma^2}{\Lambda^2}\right)+\Lambda^2\Sigma^2
-\Sigma^4\ln \left(1+\frac{\Lambda^2}{\Sigma^2}\right)\Biggr\}.
\ee
The stationarity equation for the effective potential (6) has the form
\ba
\frac{\partial V_0(\Sigma)}{\partial\Sigma}=0=
\frac{\Sigma}{4\pi^2}
\Biggl\{\frac{4\pi^2}{G}-\Lambda^2+\Sigma^2 \ln
\left (1+\frac{\Lambda^2}{\Sigma^2}\right)\Biggr\}\equiv
\frac{\Sigma}{4\pi^2}F(\Sigma).
\ea
Now one can easily see that at $G<G_c=4\pi^2/\Lambda^2$ Eq. (7) has no 
solutions apart from $\Sigma=0$. Hence, in this case fermions are massless,
and chiral invariance (2) is not broken.

If $G>G_c$, then Eq. (7) has one non-trivial solution,
$\Sigma_0(G,\Lambda) \not = 0$, such that $F(\Sigma_0)=0$. In
this case $\Sigma_0$ is a point of global minimum for the
potential $V_0(\Sigma)$. This means that spontaneous breaking of
the symmetry (2) takes place. Moreover, fermions acquire a mass
$M\equiv\Sigma_0 (G,\Lambda)$.

Let us now consider the case where $\mu>0$ and the temperature $T \not =0$. In
this case, the effective potential $V_{\mu T}(\Sigma)$ can be found
if the measure of integration in (5) is transformed in the standard 
way according to the rule \ct{dol}
$$
\int\frac{dp_0}{2\pi}\to iT\sum^{\infty}_{n=-\infty}, \qquad
p_0\to i\pi T(2n+1)+\mu.
$$
Summing there over $n$ and letting the temperature in the obtained
expression tend to zero, we obtain
\[
V_{\mu}(\Sigma)=V_0(\Sigma) -
2\int\f{d^3p}{(2\pi)^3}\theta(\mu - \sqrt{\Sigma^2+p^2})(\mu -
\sqrt{\Sigma^2+p^2}), 
\]
where $\theta(x)$ is the step function. Finally, by performing the 
momentum integration, we find
\ba
V_{\mu}(\Sigma)&=&V_0(\Sigma)-
\frac{\theta(\mu-\Sigma)}{16\pi^2}
\Biggl\{\frac{10}{3}\mu(\mu^2-\Sigma^2)^{3/2}- \Biggr.\nonumber \\
&&- ~2\mu^3\sqrt{\mu^2-\Sigma^2}+\Sigma^4\ln
\left[(\mu+\sqrt{\mu^2-\Sigma^2})^2/\Sigma^2\right]\Biggr\}.
\ea
It follows from (8) that, in the case $G<G_c$ and at arbitrary
values of the chemical potential, the chiral symmetry (2) is not broken.
However, at $G>G_c$ the model has a rich phase structure, which
is presented in Fig. 1 in terms of $\mu$ and $M$. (At $G>G_c$
one can use the fermionic mass $M$ as an independent parameter
of the theory. The three quantities $G$, $M$ and $\Lambda$ are
connected by Eq. (7).) In this figure the solid and dashed
lines represent the critical curves of the second- and
first-order phase transitions, respectively.  Furthermore, there
are two tricritical points 
\footnote{A point of the phase diagram is called a tricritical one
if, in an arbitrarily small vicinity of it, there are first- as well as second-order phase transitions.},
$\al$ and $\beta$, two massive phases
$B$ and $C$ with spontaneously broken chiral invariance as well
as the symmetric massless phase $A$ in the phase portrait of the
NJL model I (detailed calculations of the vacuum
structure of this NJL model can be found in \ct{1}).
In the present model the dynamical quark mass $\Si_0(\mu)$,
given by the global minimum of the potential $V_\mu(\Si)$
as a function of $\mu$ behaves as depicted in Fig. 2.

We should note that the phase transition from $B$ to $C$ is the
quark-matter analogue of the so-called insulator--metal phase transition in
condensed-matter physics. This is due to the fact that,
in the vacuum of phase $B$, the particle density (the analogue of
conductivity electron density in condensed materials) is zero,
while in the vacuum of phase $C$ there arises a non-zero density of
charged particles, so that it looks like a Fermi-liquid ground state of
metals.

To investigate the vacuum properties 
of the two-flavour NJL model (model II) it is again convenient to employ, 
instead of the quark Lagrangian (\ref{eq.3}), the equivalent quark--meson 
Lagrangian
\begin{eqnarray}
\label{eq.9}
\ti L_2=\bar q i \gamma^\mu \p_\mu q - 
\bar q(\sigma + i \gamma^5 \vec\tau\vec\pi)q -
 \frac{N_c}{2G}(\sigma^2+\vec\pi^2).
\end{eqnarray}
Using here calculations similar to the case with model I, one can
find the effective
potentials for $\mu=0$ and $\mu\neq 0$ expressed in terms of 
meson fields $\sigma,\vec\pi$. These potentials have the form of the
potentials (6) and (8) for the model I, respectively, with the
exception that the factor $(16\pi^2)^{-1}$ in (6),(8) has to be replaced 
by $(8\pi^2)^{-1}$. Moreover \footnote{Chiral symmetry of the two-flavour
NJL model (3) is realized in the space of meson fields
$(\sigma,\vec\pi)$ as the rotation group $O(4)$, which leaves
the ``length'' of the particle vector $(\sigma,\vec\pi)$ invariant.
 Hence, all effective potentials of this model depend
on the single variable $\Sigma=\sqrt{\sigma^2+\vec\pi^2}$. In the following 
we choose $\vec \pi=0$, assuming the absence of a pion condensate.}, in the
case under consideration $\Sigma^2=\sigma^2+\vec\pi^2$.  It
follows from this similarity that the phase
structure and the phase portrait of the model II are qualitatively the same as
those of the model I (see Fig. 1).

\section*{3. Phase structure of the model I at $\mu\neq 0$ and 
$H\neq 0$}

In the present section we shall study vacuum magnetic properties of
NJL systems. For the model I at $\mu =0$, this problem was considered in
\ct{klev,gus}.  It was shown in \ct{klev} that at $G>G_c$ the
chiral symmetry is spontaneously broken for arbitrary values of
the external magnetic field $H$, including the case $H=0$. At $G<G_c$ the
NJL model I has a symmetric vacuum at $H=0$. However, if an external
(arbitrarily small) magnetic field is switched on, then for all
$G\in (0,G_c)$ there is a spontaneous breaking of the initial $U(1)_A$
symmetry (2) \ct{gus}. This is the so-called effect of dynamical
chiral-symmetry breaking (DCSB) catalysis by an external magnetic field.
(This effect was observed for the first time in the framework of a
(2+1)-dimensional Gross--Neveu model in \ct{3} and was then
explained in \ct{gus1}. Now this effect is under intensive
investigations since it has a wide range of possible applications
in physics\footnote{Some recent references on this subject are presented
in \ct{kgk}.}.)

Let us recall some aspects of the problem at $\mu =0, H\neq 0$.
In order to find in this case  the effective
potential $V_H(\Si)$ of the NJL model I,
gauged by an external magnetic field according to
 $\partial_\mu\to D_\mu=\partial_\mu-ieA_\mu$, $A_\mu=
\delta_{\mu 2}x_1H$, one can use the well-known
proper-time method \ct{sch} or momentum-space calculations
\ct{dit}, which give the following expression:
\[
V_H(\Si)=\f{H^2}2+\f{\Si^2}{2G}+\f{eH}{8\pi^2}
\int_{0}^{\infty} \f{ds}{s^2} \exp (-s\Si^2)~\coth(eHs).
\]
In this formula $e$ has a positive 
value.
It is useful to rearrange this expression in the form
\be
\label{10}
V_H(\Si)=V_0(\Si)+Z(\Si)+\ti V_H(\Si),
\ee
where
\ba
\label{eq.11}
V_0(\Si)&=&\f{\Si^2}{2G}+\f{1}{8\pi^2}\int_{0}^{\infty}
\f{ds}{s^3} \exp (-s\Si^2),\nn \\
Z(\Si)&=&\f{H^2}2+\f{(eH)^2}{24\pi^2}
\int_{0}^{\infty} \f{ds}{s} \exp (-s\Si^2),\nn \\
\ti V_H(\Si)&=&\f{1}{8\pi^2}
\int_{0}^{\infty} \f{ds}{s^3} \exp (-s\Si^2)\left [~(eHs)\coth(eHs)
-1-\f{(eHs)^2}3~\right ],
\ea
thereby isolating the contributions of the matter, the field and the
electromagnetic interaction energy densities explicitly.
The potential $V_0(\Si)$ in (\ref{eq.11}) is up to an unimportant (infinite)
 additive constant not depending on $\Si$,
equal to expression (5). Hence, the ultra-violet (UV)-
regularized expression for it looks like (6).

The integral of the function $Z(\Si)$ is also UV-divergent, so we need to 
regularize it.
The simplest possibility is to  cut it off at the lower boundary,
which yields 
\ba
\label{eq.12}
Z(\Si)&=
&\f{H^2}2-\f{(eH)^2}{24\pi^2}\ (\ln\f{\Si^2}{\La^2}+ \gamma),
\ea
$\gamma$ being the Euler constant. Clearly,
the last term in (\ref{eq.12}) contributes to the
renormalization of the magnetic field and electric charge, in a way
similar to what occurs in quantum electrodynamics \ct{sch}.
 
The potential $\ti V_H(\Si)$ in (\ref{eq.11}) has no UV divergences, so it is
easily calculated with the help of integral tables
\ct{prud}. The final expression for $V_H(\Si)$ in terms of
 renormalized quantities is then given by
\be
\label{eq.13}
V_H(\Si)=\f{H^2}2+V_0(\Si)-\f{(eH)^2}{2\pi^2}\Bigl\{\zeta '(-1,x)-\f 12[x^2-x]\ln x
+\f{x^2}4\Bigr\},
\ee
where $x=\Si^2/(2eH)$, $\zeta (\nu,x)$ is the generalized Riemann
zeta-function and $\zeta'(-1,x)$$~=\\
d\zeta(\nu,x)/d\nu|_{\nu=-1}$.
 The global minimum point of this function is
the solution of the stationarity equation
\be
\label{eq.14}
\f {\p}{\p\Si}V_H(\Si)=\f{\Si}{4\pi^2}\{F(\Si)-I(\Si)\}=0,
\ee
where $F(\Si)$ is given in (7), and 
\ba
\label{eq.15}
I(\Si)&=&2eH\Bigl\{\ln\Ga(x)-\f 12\ln(2\pi)+x-\f 12(2x-1)\ln x\Bigr\}\nn \\
&=& \int_{0}^{\infty} \f{ds}{s^2} \exp (-s\Si^2)[~eHs\coth(eHs)-1].
\ea
One can easily see that there exists, for arbitrary fixed values of $H,G$,
only one non-trivial solution $\Si_0(H)$ of Eq.
(\ref{eq.14}), which is the global minimum point of $V_H(\Si)$.
There, $\Si_0(H)$ is a monotonically increasing function of $H$,
and at $H\to\infty$ 
\be
\label{eq.16}
\Si_0(H)\approx\f {eH}\pi \sqrt{\f G{12}}.
\ee
However, at $H\to 0$
\be
\label{eq.17}
\Si_0^2(H)\approx\cases{\f {eH}\pi\exp\{-\f 1{eH}(\f {4\pi^2}G-\La^2)\},& 
if $G<G_c$, \cr
M^2,& if $G>G_c$. \cr}
\ee

So, at $G<G_c$ and $H=0$ the NJL vacuum is chirally symmetric, but an
arbitrarily small value of the external magnetic field $H$ induces DCSB, and 
fermions acquire a non-zero mass $\Si_0(H)$ (the magnetic 
catalysis effect of DCSB).

Now let us consider the more general case, when $H\neq 0$ and $\mu\not = 0$.
In one of our previous papers \ct{7} an effective potential of
a 3D Gross--Neveu model at non-zero $H$, $\mu$ and $T$ was
obtained. Similarly, one can find the effective potential in the
NJL model I at $H,T,\mu\neq 0$:
\be
\label{eq.18}
V_{H\mu T}(\Si)=V_H(\Si)-\f{TeH}{4\pi^2}\sum_{k=0}^{\infty} \al_k
\int_{-\infty}^\infty dp\ln\left\{\left [1+\exp^{-\beta(\ve_k+\mu)}\right ]
\left [1+\exp^{-\beta(\ve_k-\mu)}\right ]\right\},
\ee
where  $\beta=1/T$, $\al_k=2-\del_{0k}$,
$\ve_k=\sqrt{\Si^2+p^2+2eHk}$,
 with $k=0,1,2, ...$ denoting Landau levels,
 and the function $V_H(\Si)$ is
given in (\ref{eq.13}). If we let the temperature in (\ref{eq.18}) tend
to zero, we obtain the effective potential of the NJL model I at
$H,\mu\neq 0$:
\be
\label{eq.19}
V_{H\mu}(\Si)=V_H(\Si)-\f{eH}{4\pi^2}\sum_{k=0}^{\infty} \al_k
\int_{-\infty}^\infty dp (\mu-\ve_k)\theta (\mu-\ve_k),
\ee
which, by performing the integration, can easily be cast into the form
\be
\label{eq.20}
V_{H\mu}(\Si)=V_H(\Si)-\f{eH}{4\pi^2}\sum_{k=0}^{\infty} \al_k
\theta (\mu-s_k)\Biggl\{\mu\sqrt{\mu^2-s_k^2}-s_k^2\ln\left
[\f {\mu+\sqrt{\mu^2-s_k^2}}{s_k}\right ]\Biggr\},
\ee
where $s_k=\sqrt{\Si^2+2eHk}$. Finally, let us present the stationarity 
equation for the potential (\ref{eq.20}):
\be
\label{eq.21}
\f {\p}{\p\Si}V_{H\mu}(\Si)\equiv\f {\Si}{4\pi^2}\phi(\Si)
=\f{\Si}{4\pi^2}\Biggl\{F(\Si)-I(\Si)
+2eH\sum_{k=0}^{\infty} \al_k \theta (\mu-s_k)\ln\left
[\f {\mu+\sqrt{\mu^2-s_k^2}}{s_k}\right ]\Biggr\}=0.
\ee
In order to get a phase portrait of the model under
consideration we should find a one-to-one correspondence
between points of the $(\mu,H)$-plane and the global minimum points of
the function (\ref{eq.20}), i.e. by solving Eq.
(\ref{eq.21}) we should find the global minimum $\Si (\mu,H)$ of the potential
(\ref{eq.20}) and then study its properties as a function of $(\mu,H)$.

\subsubsection*{3.1 The case $G<G_c$. Magnetic catalysis and chemical
potential}

In order to greatly simplify this problem, let us divide the $(\mu,H)$-plane
into a set of regions $\om_k$:
\be
\label{eq.22}
(\mu, H)=\bigcup\limits^{\infty}_{k=0}\omega_k, \qquad
\omega_k=\{(\mu, H):2eHk\leq \mu^2\leq 2eH (k+1)\}.
\ee
In the $\om_0$ region only the  first term from the series in 
(\ref{eq.20}) and (\ref{eq.21}) is non-vanishing.
So, one can find that, for the points $(\mu,H)\in\om_0$, which are above
the line $l$=$\{(\mu,H):\mu=\Si_0(H)\}$, the global minimum is at the point
$\Si=0$. Just under the curve $l$ the point $\Si=\Si_0(H)$ is a local minimum
of the potential (\ref{eq.20}), whereas $\Si=\Si_0(H)$ becomes a global
minimum, when $(\mu,H)$ lies under the curve 
$\mu=\mu_c(H)$, which is defined by the following equation:
\be
\label{eq.23}
V_{H\mu}(0)=V_{H\mu}(\Si_0(H)).
\ee
Evidently, the line $\mu=\mu_c(H)$ is the critical curve of first-order
phase transitions.
In the $\om_0$ region Eq. (\ref{eq.23}) is easily solved: 
\be
\label{eq.24}
\mu_c(H)=\f {2\pi}{\sqrt{eH}} [V_H(0)-V_H(\Si_0(H))]^{1/2}.
\ee
Using the asymptotics (\ref{eq.17}) of the solution $\Si_0(H)$ at $H\ra 0$,
we find the following behaviour of $\mu_c(H)$ at $H\to 0$:
\[
\mu_c(H)\approx \sqrt{\f {eH}{2\pi}}\exp\Bigl\{-\f 1{2eH}\left(\f {4\pi^2}G-\La^2\right)\Bigr\}.
\]

Hence, we have shown that at $\mu>\mu_c(H)$ ($G<G_c$) there exists a massless
symmetric phase of the NJL model (numerical investigations of
(\ref{eq.20}) and (\ref{eq.21})
give us a zero global minimum point for the potential $V_{H\mu}(\Si)$
in other regions $\om_1,\om_2,...$ as well). The external magnetic
field ceases to induce the DCSB at $\mu>\mu_c(H)$
(or at sufficiently small values of the magnetic field $H<H_c(\mu)$, where
$H_c(\mu)$ is the inverse function of $\mu_c(H)$).
However, under the critical curve (\ref{eq.24}) (or at $H>H_c(\mu)$), 
owing to the presence of an external magnetic field, the chiral symmetry is
spontaneously broken. Here the magnetic field induces a dynamical
fermion mass $\Si_0(H)$, which has a  $\mu$-independent value.

Lastly, we should also remark that in the NJL model (1)
the magnetic catalysis effect takes place only in the
phase with zero particle density, i.e. at $\mu<\mu_c(H)$. If
$\mu>\mu_c(H)$, we have the symmetric phase with non-zero particle
density, but here the magnetic field cannot induce DCSB.

\subsubsection*{3.2 The case $G<G_c$. Infinite cascade of massless phases}

In the previous subsection we have shown that the points $(\mu,H)$,
lying above the critical curve $\mu=\mu_c(H)$, correspond to the 
chirally symmetric ground state of the NJL model. Fermionic excitations
of this vacuum have zero masses. At first sight, it might seem that the 
properties of this 
symmetric vacuum are slightly varied, when parameters $\mu$ and $H$ are
changed. However, this is not the case, and in the region $\mu>\mu_c(H)$ 
we have infinitely many massless symmetric phases of the theory
corresponding to infinitely many Landau levels,
as well as a variety of critical curves of second-order
phase transitions. Let us next prove this.

It is well known that the state of thermodynamic equilibrium 
(the ground state) of an arbitrary quantum system is described by the 
thermodynamic potential (TDP) $\Om$, which is just the
value of the effective potential at its global minimum point.
In the case under consideration, the TDP $\Om(\mu,H)$ 
at $\mu>\mu_c(H)$ has the form
\begin{eqnarray}
\Om(\mu,H)&\equiv&V_{H\mu}(0)=V_H(0)-  \nonumber \\
&&- \frac{eH}{4\pi^2}\sum^{\infty}_{k=0}\alpha_k\theta
(\mu-\e_k)\Bigl\{\mu\sqrt{\mu^2-\e_k^2}-\e_k^2\ln \left[\left(\sqrt{\mu^2-\e_k^2}
+\mu\right)/\e_k\right]\Bigr\},
\label{eq.25}
\end{eqnarray}
where $\e_k=\sqrt{2eHk}$. We shall use the following criterion
of phase transitions: if at least one first (second) partial
derivative of $\Om(\mu,H)$ is a discontinuous function at some
point, then this is a point of a first- (second-) order
phase transition.

Using this criterion, let us show that boundaries of $\om_k$
regions (\ref{eq.22}), i.e. lines
$l_k=\{(\mu,H):\mu=\sqrt{2eHk}\}$ ($k=1,2,...$), are critical
lines of second-order phase transitions.  In an arbitrary 
$\om_k$ region the TDP (\ref{eq.25}) has the form:
\begin{eqnarray}
\Om(\mu,H)\big |_{\om_k}&\equiv&\Om_k=V_H(0)- \nonumber \\
&&- \frac{eH}{4\pi^2}\sum^{k}_{i=0}\alpha_i \theta (\mu-\epsilon_i)
\Biggl\{\mu\sqrt{\mu^2-\e_i^2}-\e_i^2\ln\left
[\f {(\sqrt{\mu^2-\e_i^2}+\mu)}{\e_i}\right ]\Biggr\}.
\label{eq.26}
\end{eqnarray}
From (\ref{eq.26}) one easily finds
\be
\label{eq.27}
\f {\p\Om_k}{\p\mu}\bigg |_{(\mu,H)\to l_{k+}}-
\f {\p\Om_{k-1}}{\p\mu}\bigg |_{(\mu,H)\to l_{k-}}=0,
\ee
as well as:
\be
\label{eq.28}
\f {\p^2\Om_k}{(\p\mu)^2}\bigg |_{(\mu,H)\to l_{k+}}-
\f {\p^2\Om_{k-1}}{(\p\mu)^2}\bigg |_{(\mu,H)\to l_{k-}}=
-\f {eH\mu}{2\pi^2\sqrt{\mu^2-\e_k^2}}\bigg |_{\mu\to \e_{k+}}\ra -\infty.
\ee
Equation (\ref{eq.27}) means that the first derivative $\p\Om/\p\mu$
is a continuous function on all lines $l_k$. However, the second
derivative $\p^2\Om/(\p\mu)^2$ has an infinite jump on each line
$l_k$ (see (\ref{eq.28})), so these lines are critical curves 
of second-order phase transitions.  (Similarly, we can prove the
discontinuity of $\p^2\Om/(\p H)^2$ and $\p^2\Om/\p\mu\p H$ on
all lines $l_n$.)

The results of the above investigations are presented in Fig. 3, where the
 phase portrait of the NJL model I at $G<G_c$ 
in the $(\mu,H)$-plane is displayed.

\subsubsection*{3.3 The case $G>G_c$}

Concerning supercritical values of the coupling constant, we shall
consider here only the case $G_c<G<(1.225...)G_c$, where the phase
portrait of the model I is qualitatively represented in Fig. 4. In
this figure one can see infinite sets of symmetric massless
$A_0,A_1,...$ phases, as well as massive phases
$C_0,C_1,...$ with DCSB. In addition, there is another massive phase $B$.
Dashed and solid lines in Fig. 4 are critical curves of first-
and second-order phase transitions, respectively.  One can also see
on this portrait infinitely many tricritical points
$t_k,s_k$ ($k=0,1,2,...$).
For a fixed value of $k$ the point $t_k$ lies
inside, but the point $s_k$ is on the left boundary of the
corresponding $\om_k$ region (\ref{eq.22}).  Each critical line
$l_k$ coincides with a part of the $\om_k$ boundary.  In Table 1, we give the 
values of the external magnetic field corresponding to
tricritical points $t_0$ and $s_0$.

The presence of an infinite cascade of massless $A_k$-phases in
the case $G>G_c$ may be proved in a way similar to what was done
in the previous subsection.  However, now an infinite set of massive
chirally non-symmetric phases is available thanks to the particular
structure of the function $\phi(\Si)$ in (\ref{eq.21}).  A detailed
numerical investigation of this function shows that, for some
values of $(\mu,H)$ inside the $C_k$ region (see Fig. 4), $\phi(\Si)$
as a function of $\Si$ qualitatively behaves like the curve,
drawn in the Fig. 5. At these values of $(\mu,H)$ there is only
one non-trivial solution $\Si_k(\mu,H)$ of the stationary
equation (\ref{eq.21}), which is the global minimum point of the
effective potential (\ref{eq.20}) and at the same time is the
quark mass in the phase $C_k$ of the theory. Remark that in
each phase $C_k$ the quark mass is a $\mu$-dependent function.
In contrast, in the phase $B$, the global minimum point is equal to
$\Si_0(H)$ (see (\ref{eq.16}), (\ref{eq.17})), which is a
$\mu$-independent quantity. Hence, the particle density
$n\equiv -\p\Om/\p\mu$ in the ground state of phase $B$ is
identically equal to zero,
 whereas in each phase $C_k$ this quantity differs from zero.  
This conclusion follows from the
definition of the thermodynamic potential $\Om$ given in the
previous subsection , as well as from the fact that $\Om$
has no $\mu$-dependence in the phase $B$.\footnote{
 Since for all points of region
$B$ we have $\mu<\Si_0(H)$, it follows from (\ref{eq.20}) that
$\Om$ = $V_{H\mu}(\Si_0(H))$ = $V_{H}(\Si_0(H))$, i.e. $\Om$ has indeed 
no $\mu$-dependence.}

If $\mu$ increases, the curve of Fig. 5 moves up and to the
right side of this figure. So, for some values of $\mu$ the
function $\phi(\Si)$ in (\ref{eq.21}) will have three zeros (see
Fig. 6): the one on the left-hand side in this figure, $\Sigma_{k+1}$,
 is a local minimum of the function $V_{H\mu}(\Si)$ , the one at the right-hand side,  $\Sigma_k$ $(\Sigma_k > \Sigma_{k+1})$, 
a global minimum of that function. 
If the chemical potential persists to grow, then at some 
critical value of $\mu$
the global minimum jumps from $\Si_k(\mu,H)$ to $\Si_{k+1}(\mu,H)$.
At this moment we have a first-order phase transition from the
massive phase $C_k$ to the massive $C_{k+1}$ one. In Fig. 4 the point of this 
phase transition lies on the curve
 $\overbrace{Mt_{k+1}}$, which is the boundary 
between regions $C_k$ and $C_{k+1}$. Hence, all lines $\overbrace{
Mt_k}$ in Fig. 4
($k=0,1,2,...$) are first-order phase-transition curves. Here we should also 
remark that all points of the line $\overbrace{t_0\mu_c(H)}$ in this 
figure are
described by Eq. (\ref{eq.23}).
Since the phase structure of the model I is so complicated, the 
dynamical quark mass $\Si(\mu,H)$, which is given by the global
minimum of the potential $V_{H\mu}(\Si)$, also has a
rather complicated $\mu,H$-dependence. For illustration, 
in Fig. 7 the schematic
behaviour of $\Si(\mu,H)$ versus $\mu$ is presented at some fixed value of
the external magnetic field $H$.

From standard textbooks on statistical physics (see, e.g.
\ct{lan}) we know that more than three curves of first-order
phase transitions (1OPT) should not intersect at one point of a
phase diagram; not more than three phases are thus allowed to
coexist in nature. However, in Fig. 4 one can see that infinitely
many curves of 1OPT cross at the one point $M$. Indeed, this
visible contradiction with the above-mentioned statement is only fictitious,
 because at the point $M$ we have a second-order phase transition
\ct{1}.  Hence, $M$ does not belong to any of the critical curves of
1OPT $\overbrace{Mt_k}$ and is therefore not a point of phase coexistence.

\section*{4. Magnetic oscillations in the model I}

Now we want to show that there arise, from the presence of infinite sets
of massless $A_k$ phases as well as of massive $C_k$ ones,
magnetic oscillations (the so-called van Alphen--de Haas-type effect) of some
physical parameters in the model I gauged by an external magnetic field.

\subsubsection*{4.1 The case $G<G_c$}

Let the chemical potential be fixed, i.e.
 $\mu = \rm {const} > \mu_{c}(H)$.
 Then on
the plane $(\mu,H)$ (see Fig. 3) we have a line that crosses critical lines
$l_1,l_2,...$ at points $H_1,H_2,...$  The
particle density $n$ and the magnetization $m$ of any
thermodynamic system are defined by the TDP in the following way:
$n=-\p\Om/\p\mu$, $m=-\p\Om/\p H$.  At $\mu$ = const these
quantities are continuous functions of the external magnetic field
only, i.e. $n\equiv n(H),~m\equiv m(H)$. We know that all the second
derivatives of $\Om(\mu,H)$ are discontinuous on every critical
line $l_n$. The functions $n(H)$ and $m(H)$, being continuous in the
interval $H\in (0,\infty)$, therefore have first derivatives that are
 discontinuous on
an infinite set of points $H_1,...,H_k,...$ Such a behaviour manifests 
itself as a phenomenon usually called oscillations. 

In condensed-matter physics \ct{haas,lan} it is a conventional rule 
to separate
the expression for a physical quantity with oscillations into two parts:
the first one is called monotonic and does not contain any oscillations,
whereas the second part, which is of particular interest here, 
contains all the
oscillations. Following this rule, we can write down, say, 
 the TDP (\ref{eq.25}) of the NJL model I in the form
\be
\label{eq.29}
\Om(\mu,H)=\Om_{mon}(\mu,H)+\Om_{osc}(\mu,H).
\ee
In order to present the oscillating part $\Om_{osc}(\mu,H)$
as well as the monotonic one $\Om_{mon}(\mu,H)$ in an
analytical form, we shall use the technique elaborated in \ct{as}, where
manifestly analytical expressions for these quantities were found 
in the case of a perfectly relativistic electron--positron gas.
This technique can be used without any difficulties in our case, too.
So, applying in (\ref{eq.25}) the Poisson summation formula \ct{lan}
\be
\label{eq.30}
\sum^{\infty}_{n=0}\alpha_n\Phi (n)
=2\sum^{\infty}_{k=0}\alpha_k
\int\limits^{\infty}_{0}\Phi (x)
\cos (2\pi kx)dx,
\ee
one can get for $\Om_{mon}(\mu,H)$
and $\Om_{osc}(\mu,H)$ the following expressions
\be
\label{eq.31}
\Omega_{mon}=V_H(0)-\f {\mu^4}{12\pi^2}-\frac{(eH)^2}{4\pi^3}
\int\limits^{\nu}_{0}dy \sum^{\infty}_{k=1}\frac{1}{k}P(\pi ky),
\ee
\be
\label{eq.32}
\Omega_{osc}=\frac{\mu}{4\pi^{3/2}}
\sum^{\infty}_{k=1}
\left (\frac{eH}{\pi k}\right)^{3/2}
[Q(\pi k \nu)\cos (\pi k\nu+\pi/4)+
P(\pi k \nu)\cos (\pi k\nu -\pi/4)],
\ee
where $\nu=\mu^2/(eH)$.
(To find (\ref{eq.31}) and (\ref{eq.32}) it is sufficient to let
the electron mass pass to zero in formula (19) of \ct{as}.)
Functions $P(x)$ and $Q(x)$ in (\ref{eq.31}) and (\ref{eq.32}) are connected with
Fresnel integrals $C(x)$ and $S(x)$ \ct{15}
\by
C(x)&=&\frac{1}{2}+\sqrt{\frac{x}{2\pi}}
[P(x)\sin x +Q(x)\cos x] \\
S(x)&=&\frac{1}{2}-\sqrt{\frac{x}{2\pi}}
[P(x)\cos x -Q (x)\sin x].
\ey
They have, at $x\to\infty$, the following asymptotics \ct{15}:
\[
P(x)=x^{-1}-\frac{3}{4}x^{-3}+...,~~~Q(x)=-\frac{1}{2}x^{-2}+
\frac{15}{8}x^{-4}+...
\]
Formula (\ref{eq.32}) presents, in a manifestly analytical form, the
oscillating part of the TDP (\ref{eq.25}) for the NJL model at
$G<G_c$. In the case under consideration, since the TDP is
proportional to the pressure of the system, one can conclude
that the pressure in the NJL model oscillates when $H\to 0$, too.  It
follows from (\ref{eq.32}) that the frequency of oscillations over the 
parameter $(eH)^{-1}$ equals $\mu^2/2$.  Then, starting
from (\ref{eq.32}), one can easily find the corresponding expressions for
the oscillating parts of $n(H)$ and $m(H)$. These quantities
oscillate at $H\to 0$ with the same frequency $\mu^2/2$ and
have a rather involved form, so we do not present them here.

Finally, we should note that the character of magnetic
oscillations in the NJL model at $G<G_c$ resembles the magnetic
oscillations in massless quantum electrodynamics \ct{per,as}.
This circumstance is conditioned by the resemblance of the vacuum
properties in the two models. Indeed, both in the NJL model and in QED,
for fixed $\mu$ and varying values of $H$, there are infinitely many
second-order phase transitions (see the appendix, where the vacuum
structure of QED at $\mu,H\neq 0$ is considered).

\subsubsection*{4.2 The case $G_c<G<(1.225...)G_c$}

Here at $\mu>\mu_{1c}(M)$ (see Fig. 4) the TDP of the NJL model as well as all 
thermodynamical parameters of the system oscillate with the 
frequency $\mu^2/2$.
This can be shown in a way similar to what was done in the previous section.  

However, at $M<\mu<\mu_{1c}(M)$, the character of magnetic oscillations 
is changed. Let us next prove this.
First of all, in this case we have another expression for the TDP 
of the system 
\be
\label{eq.33}
\Om(\mu,H)=V_{H\mu}(\Si(\mu,H)),
\ee
where $V_{H\mu}(\Si)$ is given in (\ref{eq.20}) and $\Si(\mu,H)$ 
is the non-trivial solution of the stationarity equation (\ref{eq.21}).
In each of the massive phases $C_k$ the $\mu$- and $H$-dependent function
$\Si(\mu,H)$ coincides with the corresponding fermionic mass
$\Si_k(\mu,H)$ (see subsection 3.3). Using in (\ref{eq.20}) again
the Poisson summation formula (\ref{eq.30}), one can easily select the 
oscillating part of the TDP (\ref{eq.33})
in a manifestly analytical form
\ba
\label{eq.34}
\Omega_{osc}(\mu,H)&=&\frac{\mu \theta (\mu -\Si(\mu,H))}{4\pi^{3/2}}
\sum^{\infty}_{k=1}
\left (\frac{eH}{\pi k}\right)^{3/2}
[Q(\pi k\nu)\cos (2\pi k\omega+\pi/4)+\nn \\
&&+ P(\pi k\nu)\cos (2\pi k\omega -\pi/4)],
\ea
where $\nu=\mu^2/(eH)$, $\om=(\mu^2-\Si^2(\mu,H))/(2eH)$. From
(\ref{eq.34}) one can see that as a function of the variable $(eH)^{-1}$, 
the TDP
(\ref{eq.33}) oscillates with frequency
$(\mu^2-\Si^2(\mu,H))/2$ if this variable tends to infinity. 
Since $\Om(\mu,H)$ is equal, up to a sign, to the 
pressure in the ground state of the system, also in the present case 
the pressure in the NJL model is an oscillating quantity.
Moreover, taking into account the stationarity equation (\ref{eq.21}), one
can easily derive manifest expressions
for oscillating parts of other thermodynamic quantities such as
particle density $n=-\p\Om/\p\mu$ and magnetization
$m=-\p\Om/\p H$ from (\ref{eq.33}) and (\ref{eq.34}),
\ba
\label{eq.35}
m_{osc}&=&-\f {(\mu^2-\Si^2(\mu,H))^{3/2}}{4\sqrt{2}\pi^3\mu\om^{1/2}}
\sum^{\infty}_{k=1}\f {\sin(2\pi k\om-\pi /4)}{k^{3/2}}+o\left 
(\f 1{\om^{1/2}}\right ),\nn \\
n_{osc}&=&\f {(\mu^2-\Si^2(\mu,H))^{3/2}}{4\sqrt{2}\pi^3\mu\om^{3/2}}
\sum^{\infty}_{k=1}\f {\sin(2\pi k\om-\pi /4)}{k^{3/2}}.
\ea
It is clear from (\ref{eq.35}) that particle density and magnetization in
the ground state of the NJL model oscillate with the same frequency as $\Om$.

Comparing magnetic oscillations in the NJL model I and in QED, we
see three main differences. Let us remark that in QED the frequency
of magnetic oscillations (over the variable $(eH)^{-1}$) is equal 
to $(\mu^2-M^2)/2$, where $M$
is the electron mass \ct{as}. In the NJL model, in
contrast to QED, the magnetic oscillation frequency is an
$H$-dependent quantity. So, strictly speaking, in the NJL model magnetic 
oscillations are not periodic ones. 
Similar
peculiarities of magnetic oscillations are observed in some
ferromagnetic semiconductive materials such as $\rm HgCr_2Se_4$
\ct{bal}, where non-periodic magnetic oscillations over the variable
$(eH)^{-1}$ were found to exist for electric
conductivity\footnote{Magnetic oscillations of the electric
conductivity, which is proportional to the particle density, are
known as the Shubnikov--de Haas effect \ct{haas}.} as well as magnetization.
 This is the first distinction.

A second difference is the character of the oscillations in the two models:
in QED, magnetic oscillations are accompanied by second-order
phase transitions (see Appendix), while in the NJL model they occur as a result of an infinite cascade of first-order phase transitions.

Thirdly, we should remark that in the NJL model not only thermodynamic
quantities oscillate, but some 
dynamical parameters of the system do as well. This concerns, in particular,  
oscillations of the 
dynamical quark mass. In fact, by applying the Poisson summation 
formula (\ref{eq.30})
in the stationarity equation (\ref{eq.21}) and searching for the 
solution $\Si(\mu,H)$
of this equation in the form $\Si(\mu,H)=\Si_{mon}+\Si_{osc}$, one can easily
find the following expressions for $H\to 0$:
\ba 
\label{eq.36}
\Si_{osc}&=&\f {(\mu^2-\ti M^2)^{3/2}}{\sqrt{2}\pi\mu\ti\om^{3/2}f'(\ti M)}
\sum^{\infty}_{k=1}\f {\sin(2\pi k\ti\om-\pi /4)}{k^{3/2}}+o\left 
(\f 1{\om^{3/2}}\right ),\nn \\
\Si_{mon}&=&\ti M+\f {\mu (\mu^2-\ti M^2)^{3/2}}{12\ti M^2f'(\ti M)\ti\om^2}
\left [1+\f {\sqrt{\mu^2-\ti M^2}}\mu \right ]+o\left 
(\f 1{\om^{3/2}}\right ),
\ea
where $\ti\om=(\mu^2-\ti M^2)/2(eH)$, $\ti M\equiv \Si_0(\mu)$
is the quark mass at $H=0$, $\mu\neq 0$ (see Fig. 2) and
\[
f(x)=F(x)+2\mu\sqrt{\mu^2-x^2}-2x^2\ln\f {\mu+\sqrt{\mu^2-x^2}}{x}
\]
($F(\Si)$ is defined in (7)). Hence, in the framework of the NJL model I 
the quark mass $\Si(\mu,H)$, as well as other dynamical
quantities composed from it, oscillate in the presence 
of an external magnetic field. 

\section*{5. Magnetic oscillations in the model II}

Next, let us consider magnetic oscillations in the more realistic  NJL 
model II containing two kinds of quarks: $u$- and $d$-quarks with electric
charges $e_1$ and $e_2$, respectively. The effective potential $v_{h\mu}$ in
the case under consideration is a trivial generalization of $V_{H\mu}$ 
derived in the one-flavour case
\ba
\label{eq.37}
v_{h\mu}(\Si)=-\f{H^2}2-\f {\Si^2}{2G}+\sum^2_{i=1}V_{e_iH\mu}(\Si),
\ea
where $\Si=\sqrt{\si^2+\vec\pi^2}$, and $V_{e_iH\mu}(\Si)$ is equal to
$V_{H\mu}(\Si)$ (\ref{eq.20}), with $e$ replaced by $|e_i|$.

Qualitatively, the phase structure of the NJL model II is the same 
as that of model I.  So, at $G<G_c$ we have an infinite set of
massless phases (similar to the phase portrait of the model I in
Fig. 3) reflecting the infinite set of Landau levels that is the
basis for magnetic oscillations. Using the analytical methods of
section 4, one can easily select in the present case the
oscillating part of the thermodynamic potential:
\be
\label{eq.38}
\Omega_{osc}=\frac{\mu}{4\pi^{3/2}}
\sum^2_{i=1}\sum^{\infty}_{k=1}
\left (\frac{|e_i|H}{\pi k}\right)^{3/2}
[Q(\pi k \nu_i)\cos (\pi k\nu_i+\pi/4)+
P(\pi k \nu_i)\cos (\pi k\nu_i -\pi/4)],
\ee
where $\nu_i=\mu^2/(|e_i|H)$. Hence, in the model II, in
contrast to the model I and QED, we have a superposition of
two oscillating modes. At growing values of the parameter
$(eH)^{-1}$, the frequency of oscillations  in each of the modes is
equal to $e\mu^2/(2|e_i|)$.

At $G_c<G$ there is a finite vicinity of $G_c$ in which the
phase portrait of the model II in the $(\mu,H)$-plane is similar
to the one in the model I (see Fig. 4). So, in the case under
consideration we have infinite sets of massless and massive
phases as well. The cascade of massive phases is the foundation
for non-periodic magnetic oscillations. Indeed, from
(\ref{eq.37}) it follows that the oscillating part of TDP has the form
\ba
\label{eq.39}
\Omega_{osc}(\mu,H)&=&\frac{\mu \theta (\mu -\Si(\mu,H))}{4\pi^{3/2}}
\sum^2_{i=1}\sum^{\infty}_{k=1}
\left (\frac{|e_i|H}{\pi k}\right)^{3/2}
[Q(\pi k\nu_i)\cos (2\pi k\omega_i+\pi/4)+\nn \\
&&+ P(\pi k\nu_i)\cos (2\pi k\omega_i -\pi/4)],
\ea
where $\nu_i=\mu^2/(|e_i|H)$, $\om_i=(\mu^2-\Si^2(\mu,H))/(2|e_i|H)$, and
$\Si(\mu,H)$ is the global minimum point of the effective potential (37).
As in the previous model, $\Si(\mu,H)$ in the present case is an $H$-dependent
function. This means that magnetic oscillations in the NJL model II are
composed of two non-periodic harmonics, because each of them has, 
as a function of 
the variable $(eH)^{-1}$, the $H$-dependent 
frequency $e(\mu^2-\Si^2(\mu,H))/(2|e_i|)$.

\section*{6. Summary and conclusions}

In the present paper we have studied the magnetic properties of
a many-body system of cold and dense quark matter with
four-fermion interactions. In particular, we have investigated
the ground-state (vacuum) structure of two simple NJL models
with one or two quark flavours, respectively, which are taken at
non-zero chemical potential $\mu$ and magnetic field $H$.

As it turns out, in both types of models there exists a phase
$B$ (see Figs. 3 and 4) in which the quark mass is equal to $\Si_0(H)$,
i.e. it is a $\mu$-independent quantity.  Since this phase is
achieved in the region $\mu<\Si_0(H)$, the resulting particle
density is expected to be zero, which is supported by our
calculation. Clearly, this is in agreement with the physical
interpretation of the chemical potential as the energy required
to create one particle in the system.  Indeed, the energy $\mu$,
which in the phase $B$ is smaller than the quark mass, is not
sufficient to create a particle, so that in the ground state of
this phase the particle density must vanish.

Most interestingly, we have shown that in NJL models there
exist an infinite set of massless chirally invariant phases
(phases $A_0,A_1,...$ in Figs. 3 and 4), which lead to periodic
magnetic oscillations of some thermodynamic quantities of the
system (so-called van Alphen--de Haas effect).  In NJL models,
this effect is observed at weak couplings ($G<G_c$, where
$G_c=4\pi^2/\La^2$) or at sufficiently high values of the
chemical potential and resembles magnetic oscillations in
massless QED.

Furthermore, for some finite interval of the coupling constant
$G_c<G<G_1$, where in the framework of the model I
$G_1=(1.225...)G_c$, the phase structure of NJL models I and II
contains an infinite set of massive chirally non-invariant phases
(phases $C_0,C_1,...$ in Fig. 4).  This is the basis for
non-periodic magnetic oscillations of some thermodynamic
parameters, since the dynamical quark mass in each of the phases
$C_k$ is now itself an $H$-dependent quantity. (Notice that
analogous non-periodic magnetic oscillations were recently found
to exist in some condensed-matter materials \ct{bal}, too.) We
should also remark that, at $G_c<G<G_1$, some dynamical parameters
in NJL systems, such as quark masses, oscillate over
$(eH)^{-1}$ as well. This is an unknown fact in the standard
condensed-matter theory of the van Alphen--de Haas effect.

Moreover, our numerical analysis shows that for values of the 
coupling constant $G\in (G_1,G_2)$, where
$G_2\approx 40G_c$, we have in both models a finite number of massive phases
$C_k$ (the number of massless phases is infinite as before).
 At larger values of the coupling constant, i.e. at $G>G_2$,
there exist no massive phases in the phase structure of NJL
models at all, except the trivial phase $B$.

Finally, it is interesting to note that for fixed magnetic field $H$
 the dynamical quark mass 
$\Sigma(\mu, H)$ of NJL models discontinuously jumps as a function of 
$\mu$, at points 
$\mu_0, \mu_1, \mu_2,...$ (see Fig. 7), thus reflecting the structure of the 
underlying phase portrait shown in Fig. 4.

In conclusion, we have shown that NJL models at $\mu, H\ne 0$ exhibit 
an interesting phase structure and a set of oscillating quantities,
which are richer than in the corresponding QED case. Note also that the 
above approach can be further employed for a combined study of 
quark and diquark condensates of NJL models taken at $\mu, H\ne 0$. 
Work in this direction is under way.

\vspace{0.5cm}

One of the authors (D.E.) gratefully acknowledges the kind support and warm
hospitality of the colleagues of the Theory Division at CERN.
This work was supported in part by the Russian Fund for Fundamental 
Research, project 98-02-16690, and by DFG-project 436 RUS 113/477.

\subsubsection*{\leftline{\ul{Appendix}~~~~~}}
\subsubsection*{The QED vacuum structure in the presence of $\mu$ and $H$.}

In the framework of QED and in a one-loop approximation,
 the effective Lagrangian in the presence of an external homogeneous 
magnetic field has the
following form \ct{per,as}:
$$
L_{eff}(\mu,H)=L_1(H)-\Omega_{QED}(\mu,H), \eqno(A.1)
$$
where $L_1(H)$ is the Lagrangian at $\mu =0,H\neq 0$. Since it does not 
influence the phase structure of QED, we do not present its explicit 
form here,
but refer to  \ct{per,as}.
The second part of $L_{eff}(\mu,H)$ is exactly the thermodynamic potential
of a perfect electron--positron gas
$$
\Om_{QED}(\mu,H)=-\frac{eH}{4\pi^2}\sum^{\infty}_{n=0}\alpha_n\theta
(\mu-\e_n)\Bigl\{\mu\sqrt{\mu^2-\e_n^2}-\e_n^2\ln \left[(\sqrt{\mu^2-\e_n^2}
+\mu)/\e_n\right]\Bigr\}. \eqno(A.2)
$$
In (A.2) we have introduced the notation: $\e_n=\sqrt{M^2+2eHn}$, where
$M$ is the electron mass at $H=0, \mu=0$, and $\al_n=2-\del_{0n}.$

Using the results and methods of subsection 3.2 one can now easily
show that, on each line $l_n=\{(\mu,H):\mu=\sqrt{M^2+2eHn}\}$
of the $(\mu,H)$-plane, all second derivatives of
$L_{eff}(\mu,H)$ (A.1) are discontinuous. So, the lines
$l_n~(n=0,1,...)$ in Fig. 8 are the curves of second-order
phase transitions.  They divide the $(\mu,H)$-plane into an
infinite set of regions $C_n~(n=0,1,...)$ corresponding to
different massive phases of QED 
(associated to Landau levels) with the same electron mass $M$.
Nevertheless, each phase $C_n$
is  characterized by such physical quantities as particle
density $n(\mu,H)$
and magnetization
$m(\mu,H)$. On each phase boundary $l_n$ these quantities are
continuous functions.  However, their first derivatives are
discontinuous functions on each line $l_n$, so the derivative
jump of $n$ or $m$ is the signal of a second-order phase
transition.

As in the previously considered case (see subsection 3.3 and Fig. 4) the point
$M$ in Fig. 8, where all lines $l_n$ intersect, is a special point differing 
from other points of the lines $l_n$. 
Indeed, at $H=0$ we have \ct{as}
$$
\Omega_{QED}(\mu,0)=-\frac{\theta(\mu-M)}{6\pi^2}
\int\limits^{\mu}_{M}\frac{xdx}{\sqrt{x^2-M^2}}
(\mu-x)^2(\mu+2x). \eqno(A.3)
$$
It follows from (A.3) that only the third derivative
$\p^3\Om_{QED}(\mu,0)/(\p\mu)^3$ is a discontinuous function at
the point $\mu=M$.  At the same time, in other points of critical
curves $l_n$, already the second derivative
$\p^2\Om_{QED}(\mu,H)/(\p\mu)^2$ is discontinuous.

Here we have considered only the relativistic case, but one can easily
show that the non-relativistic electron gas at $\mu,H\neq 0$ has an 
infinite cascade of massive phases, too. So, at the basis of the van Alphen--de Haas and Shubnikov--de Haas effects lies an infinite set of
second-order phase transitions.

\vspace*{0.5cm}


\newpage
\begin{center}
Figure captions
\end{center}
\vspace{0.5cm}

Fig. 1 Phase portrait of the NJL model at non-zero $\mu$ and
for arbitrary values of the fermion mass $M$. Phases $B$ and $C$ are 
massive and non-symmetric, $A$ is a chirally symmetric phase. 
Here $\mu_{2c}(M)=M$,
$\mu_{1c}(M)=\sqrt{\f 12 M^2\ln(1+\La^2/M^2)}$, 
$\mu_{3c}(M)$ is the solution of the equation $V_\mu(0)=V_\mu(M)$, 
$M_{2c}=\La/(2.21...)$,
$M_{1c}$ is the solution of the equation $\mu_{1c}^2(M_{1c})=\La^2/(4e)$.
In phase $B$ the particle density in the ground state is equal to zero,
whereas in phase $C$ it is non zero.
 Solid and dashed lines represent critical curves of second- and 
first-order phase transitions, respectively; $\alpha$ and
$\beta$ denote tricritical points.
\vspace{0.5cm}

Fig. 2 The behaviour of the dynamical quark mass $\Si_0(\mu)$ as a function of
 $\mu$
for the case $M<M_{1c}$ and $H=0$ within the framework of the 
model I.
\vspace{0.5cm}

Fig. 3 Phase portrait of the gauged model I at $G<G_c$.
Solid lines $l_k$ are given by $l_k=\{(\mu,H):\mu=\sqrt{2eHk}\}$.
They are critical curves of second-order phase transitions. 
The dashed line of first-order phase transitions is defined by Eq. (23).
\vspace{0.5cm}

Fig. 4 Phase portrait of the gauged model I at $G_c<G<(1.225...)G_c$.
Here $M$ is the quark mass at $\mu =0,H=0$, and the quantity $\mu_{1c}(M)$
is presented in Fig. 1. The dashed line $\overbrace{t_0\mu_c(H)}$ 
is defined by Eq. (23).
In this case one has infinite sets of symmetric massless phases 
$A_0, A_1, ...$
as well as massive phases 
$C_0, C_1, ...$ with DCSB. In addition there exists another massive phase $B$.
\vspace{0.5cm}

Fig. 5 Typical behaviour of $\phi(\Si)$ (21) for some points $(\mu,H)\in C_k$. Here $\si_n=\\
  \sqrt{\mu^2-2eHn}$.
\vspace{0.5cm}

Fig. 6 For increasing $\mu$, 
there arise three zeroes of the 
function $\phi (\Si)$ (21) defining two local minima 
$\Si_k$ and $\Si_{k+1}$($\Si_{k+1} < \Si_k$) of the effective
potential, respectively. The global minimum 
of $V_{H\mu}(\Si)$ lies in one of them and passes by a jump from one local
minimum to another one depending on the values of $\mu$.
\vspace{0.5cm}

Fig. 7 Schematic representation of the dynamical quark mass $\Si(\mu,H)$ as
function of $\mu$ for fixed magnetic field  qualitatively reflecting
the structure of the phase portrait in Fig. 4. Here the magnetic field is 
fixed 
in the interval $(H_{s_2},H_{t_2})$, where $s_2,t_2$ are tricritical points
(see Fig. 4). The $\mu_k$,$\ti\mu_c(H)$ $(k=0,1,2)$ are the values of the
chemical potential at which the line $H$ = const crosses in Fig. 4 the
critical curves $\overbrace{Mt_k}$ and $\overbrace{s_2t_2}$, respectively.
\vspace{0.5cm}

Fig. 8 Phase portrait of QED at $\mu,H\neq 0$. All the lines $l_n$ are
the curves of second-order phase transitions.

\begin{center}
Table caption
\end{center}
\vspace{0.5cm}

Table 1. Values of the external magnetic field corresponding to
tricritical points $t_0$ and $s_0$ (see Fig. 4) for different ratios
of coupling constants $G/G_c$.
\newpage
\unitlength=1mm
\special{em:linewidth 0.4pt}
\linethickness{0.4pt}
\begin{picture}(145.67,88.33)
\put(11.00,8.33){\vector(0,1){80.00}}
\put(4.67,17.33){\vector(1,0){141.00}}
\linethickness{1.0pt}
\put(11.00,17.33){\line(2,1){80.33}}
\put(59.67,63.00){\line(5,1){7.67}}
\put(71.67,64.67){\line(5,-1){8.67}}
\put(83.33,61.33){\line(2,-1){10.00}}
\put(95.67,55.00){\line(4,-1){9.67}}
\put(108.67,52.00){\line(6,1){11.00}}
\put(122.00,54.67){\line(3,2){8.00}}
\bezier{336}(11.00,17.33)(26.67,65.00)(60.00,63.00)
\put(60.00,17.33){\line(0,1){2.33}}
\put(91.00,17.33){\line(0,1){2.33}}
\put(140.33,21.33){\makebox(0,0)[cc]{$\scs{M}$}}
\put(16.67,85.33){\makebox(0,0)[cc]{$\mu$}}
\put(84.33,37.67){\makebox(0,0)[cc]{$B$}}
\put(44.33,47.67){\makebox(0,0)[cc]{$C$}}
\put(60.00,63.00){\circle*{1.33}}
\put(60.00,68.00){\makebox(0,0)[cc]{$\alpha$}}
\put(92.00,61.67){\makebox(0,0)[cc]{$\beta$}}
\put(91.00,57.00){\circle*{1.33}}
\put(37.67,80.00){\makebox(0,0)[cb]{$A$}}
\put(21.67,50.67){\makebox(0,0)[cc]{$\scriptsize{\mu_{1c}(M)}$}}
\put(53.00,33.00){\makebox(0,0)[cc]{$\scriptsize{\mu_{2c}(M)}$}}
\put(124.67,62.00){\makebox(0,0)[cc]{$\scriptsize{\mu_{3c}(M)}$}}
\put(60.00,13.00){\makebox(0,0)[cc]{$M_{1c}$}}
\put(91.00,13.00){\makebox(0,0)[cc]{$M_{2c}$}}
\put(70.33,5.33){\makebox(0,0)[cc]{Fig. 1}}
\end{picture}
\newpage
\unitlength=1mm
\special{em:linewidth 0.4pt}
\linethickness{0.4pt}
\begin{picture}(131.00,107.33)
\put(11.00,26.67){\vector(1,0){120.00}}
\put(19.00,18.33){\vector(0,1){89.00}}
\linethickness{0.1pt}
\put(61.67,26.67){\line(0,1){2.00}}
\put(125.67,32.67){\makebox(0,0)[cc]{$\mu$}}
\put(12.33,100.33){\makebox(0,0)[cc]{$\Si_0(\mu)$}}
\put(12.00,68.67){\makebox(0,0)[cc]{$\scriptsize{M}$}}
\put(61.67,23.00){\makebox(0,0)[cc]{$\scriptsize{M}$}}
\put(85.67,23.00){\makebox(0,0)[cc]{$\scriptsize{\mu_{1c}(M)}$}}
\put(72.67,12.33){\makebox(0,0)[cc]{Fig. 2}}
\linethickness{1pt}
\put(19.00,68.67){\line(1,0){42.67}}
\bezier{264}(62.00,68.67)(86.00,68.67)(85.67,26.67)
\end{picture}
\newpage
\unitlength=1mm
\special{em:linewidth 0.4pt}
\linethickness{0.4pt}
\begin{picture}(140.33,138.67)
\put(13.00,31.33){\vector(0,1){107.33}}
\put(7.33,39.00){\vector(1,0){133.00}}
\put(13.00,39.00){\line(1,1){70.00}}
\put(13.00,39.00){\line(3,5){45.33}}
\put(13.00,39.00){\line(1,6){13.67}}
\put(13.00,39.00){\line(6,1){8.00}}
\put(23.00,40.67){\line(5,1){8.67}}
\put(34.33,43.00){\line(4,1){8.33}}
\put(45.33,45.33){\line(3,1){8.67}}
\put(56.67,49.00){\line(5,2){8.33}}
\put(67.00,53.67){\line(3,2){7.67}}
\put(76.33,60.00){\line(6,5){6.67}}
\put(84.67,67.33){\line(1,1){6.33}}
\put(92.67,76.00){\line(5,6){5.67}}
\put(100.00,84.67){\line(3,4){5.67}}
\put(66.00,30.33){\makebox(0,0)[cc]{Fig. 3}}
\put(132.33,43.67){\makebox(0,0)[cc]{$\scs{\sqrt{2eH}}$}}
\put(7.67,134.67){\makebox(0,0)[cc]{$\mu$}}
\put(103.33,54.33){\makebox(0,0)[cc]{$B$}}
\put(69.00,73.67){\makebox(0,0)[cc]{$A_0$}}
\put(53.00,89.33){\makebox(0,0)[cc]{$A_1$}}
\put(107.00,95.67){\makebox(0,0)[cc]{$\scs{\mu_c(H)}$}}
\put(84.67,112.00){\makebox(0,0)[cc]{$l_1$}}
\put(60.33,117.67){\makebox(0,0)[cc]{$l_2$}}
\put(30.33,124.67){\makebox(0,0)[cc]{$l_k$}}
\put(41.67,96.67){\circle*{0.67}}
\put(35.00,98.33){\circle*{0.67}}
\put(29.00,99.67){\circle*{0.67}}
\put(20.00,101.00){\circle*{0.67}}
\put(16.00,101.67){\circle*{0.67}}
\end{picture}
\newpage
\unitlength=1.00mm
\special{em:linewidth 0.4pt}
\linethickness{0.4pt}
\begin{picture}(147.00,132.67)
\put(4.67,15.00){\vector(1,0){136.67}}
\put(14.33,7.67){\vector(0,1){125.00}}
\put(14.33,23.67){\circle*{1.33}}
\put(14.33,23.67){\line(1,0){6.67}}
\put(23.33,23.67){\line(6,1){8.00}}
\put(33.33,24.67){\line(6,1){8.33}}
\put(44.33,26.67){\line(6,1){8.00}}
\put(55.00,28.67){\line(4,1){9.00}}
\put(66.67,31.33){\line(4,1){8.33}}
\put(77.33,33.67){\line(4,1){8.33}}
\put(88.00,36.00){\line(2,1){7.33}}
\put(97.33,40.33){\line(5,3){7.33}}
\put(106.67,45.67){\line(2,1){6.67}}
\put(116.00,50.00){\line(4,1){8.33}}
\put(126.67,53.33){\line(-1,-2){0.33}}
\put(126.33,52.67){\line(6,1){7.00}}
\put(136.00,54.33){\line(6,1){7.33}}
\put(107.00,45.67){\circle*{1.00}}
\put(93.67,67.33){\circle*{1.00}}
\put(81.67,69.67){\circle*{1.00}}
\put(68.33,67.67){\circle*{1.00}}
\put(57.67,70.67){\circle*{1.00}}
\put(46.33,69.67){\circle*{1.00}}
\put(37.33,69.33){\circle*{1.00}}
\put(30.33,70.67){\circle*{1.00}}
\put(23.33,72.33){\circle*{0.00}}
\put(19.33,71.67){\circle*{0.00}}
\put(16.33,70.67){\circle*{0.00}}
\put(14.33,70.33){\circle*{1.33}}
\put(93.67,67.33){\line(-1,2){2.00}}
\put(81.67,69.67){\line(-4,-5){4.67}}
\put(75.67,62.00){\line(-6,-5){6.00}}
\put(68.33,55.33){\line(-6,-5){6.33}}
\put(60.00,48.00){\line(-3,-2){6.67}}
\put(51.00,42.00){\line(-5,-3){7.67}}
\put(40.67,35.33){\line(-2,-1){8.00}}
\put(30.67,29.67){\line(-5,-2){6.67}}
\put(21.33,25.67){\line(-4,-1){7.00}}
\put(68.33,67.67){\line(-2,5){1.67}}
\put(57.67,70.33){\line(-2,-3){3.33}}
\put(53.33,63.33){\line(-3,-5){3.67}}
\put(48.00,55.67){\line(-3,-4){4.33}}
\put(42.00,47.67){\line(-4,-5){4.67}}
\put(35.67,40.00){\line(-1,-1){4.67}}
\put(29.00,33.67){\line(-5,-4){6.33}}
\put(21.00,27.67){\line(-5,-3){6.67}}
\put(37.33,69.33){\line(-2,-5){3.00}}
\put(33.67,60.33){\line(-1,-2){3.67}}
\put(28.67,50.33){\line(-2,-5){2.67}}
\put(24.67,41.33){\line(-1,-2){4.00}}
\put(19.67,31.00){\line(-3,-4){5.67}}
\put(93.67,67.33){\line(5,3){41.33}}
\put(68.33,67.33){\line(1,1){44.33}}
\put(46.33,69.33){\line(3,5){27.33}}
\put(30.33,70.67){\line(1,4){11.33}}
\put(147.00,15.00){\makebox(0,0)[cc]{$\sqrt{eH}$}}
\put(8.33,127.00){\makebox(0,0)[cc]{$\mu$}}
\put(8.33,69.67){\makebox(0,0)[cc]{$\scriptsize{\mu_{1c}(M)}$}}
\put(9.33,23.67){\makebox(0,0)[cc]{$\scriptsize{M}$}}
\put(95.00,25.33){\makebox(0,0)[cc]{$B$}}
\put(73.67,42.67){\makebox(0,0)[cc]{$C_0$}}
\put(51.33,49.00){\makebox(0,0)[cc]{$C_1$}}
\put(36.00,50.00){\makebox(0,0)[cc]{$C_2$}}
\put(23.33,51.00){\makebox(0,0)[cc]{$C_3$}}
\put(19.67,51.33){\circle*{0.00}}
\put(17.00,51.00){\circle*{0.00}}
\put(124.33,72.00){\makebox(0,0)[cc]{$A_0$}}
\put(101.67,84.67){\makebox(0,0)[cc]{$A_1$}}
\put(75.33,93.33){\makebox(0,0)[cc]{$A_2$}}
\put(75.33,6.00){\makebox(0,0)[cc]{Fig. 4}}
\put(48.00,95.33){\makebox(0,0)[cc]{$A_3$}}
\put(28.33,95.00){\circle*{0.00}}
\put(23.33,95.00){\circle*{0.00}}
\put(18.00,95.00){\circle*{0.00}}
\put(134.67,49.67){\makebox(0,0)[cc]{$\mu_c(H)$}}
\put(133.33,94.33){\makebox(0,0)[cc]{$l_1$}}
\put(109.67,114.33){\makebox(0,0)[cc]{$l_2$}}
\put(74.67,117.67){\makebox(0,0)[cc]{$l_3$}}
\put(42.67,120.33){\makebox(0,0)[cc]{$l_4$}}
\put(109.33,43.33){\makebox(0,0)[cc]{$t_0$}}
\put(90.33,64.67){\makebox(0,0)[cc]{$s_0$}}
\put(81.00,73.00){\makebox(0,0)[cc]{$t_1$}}
\put(65.67,65.33){\makebox(0,0)[cc]{$s_1$}}
\put(57.33,73.67){\makebox(0,0)[cc]{$t_2$}}
\put(44.33,66.67){\makebox(0,0)[cc]{$s_2$}}
\put(37.00,73.00){\makebox(0,0)[cc]{$t_3$}}
\put(28.00,67.67){\makebox(0,0)[cc]{$s_3$}}
\put(57.67,70.67){\line(5,6){3.00}}
\put(30.33,70.67){\line(-5,2){4.33}}
\linethickness{1.0pt}
\bezier{116}(107.00,45.67)(96.00,51.33)(93.67,67.33)
\bezier{84}(81.67,69.33)(69.00,61.33)(68.33,67.33)
\bezier{72}(57.67,70.33)(47.67,63.67)(46.33,69.33)
\bezier{52}(37.67,69.00)(31.00,65.00)(30.33,70.33)
\bezier{12}(37.33,69.33)(38.67,71.00)(40.00,72.00)
\bezier{12}(46.33,69.67)(45.67,71.33)(45.00,72.67)
\bezier{12}(41.00,73.00)(42.00,74.33)(43.67,73.67)
\bezier{28}(61.67,74.67)(63.67,76.67)(66.33,72.67)
\bezier{16}(81.67,69.33)(83.33,71.33)(84.33,72.33)
\bezier{28}(85.67,73.33)(88.67,75.00)(90.67,72.67)
\end{picture}
\newpage
\unitlength=1mm
\special{em:linewidth 0.4pt}
\linethickness{0.4pt}
\begin{picture}(149.33,140.00)
\put(2.33,140.00){\line(1,0){147.00}}
\put(2.33,125.00){\line(1,0){147.00}}
\put(2.33,110.00){\line(1,0){147.00}}
\put(2.33,95.00){\line(1,0){147.00}}
\put(2.33,140.00){\line(0,-1){45.00}}
\put(149.33,140.00){\line(0,-1){45.00}}
\put(35.00,140.00){\line(0,-1){45.00}}
\put(88.00,140.00){\line(0,-1){45.00}}
\put(61.33,140.00){\line(0,-1){45.00}}
\put(118.00,140.00){\line(0,-1){45.00}}
\put(14.67,132.00){\makebox(0,0)[cc]{$G/G_c$}}
\put(15.00,117.00){\makebox(0,0)[cc]{$eH_{t_0}/\Lambda^2$}}
\put(15.33,102.00){\makebox(0,0)[cc]{$eH_{s_0}/\Lambda^2$}}
\put(76.67,83.33){\makebox(0,0)[cc]{Table 1}}
\put(47.67,132.33){\makebox(0,0)[cc]{1.01}}
\put(74.67,131.67){\makebox(0,0)[cc]{1.1}}
\put(103.00,131.67){\makebox(0,0)[cc]{1.15}}
\put(133.33,132.00){\makebox(0,0)[cc]{1.2}}
\put(47.33,116.67){\makebox(0,0)[cc]{0.0129...}}
\put(74.33,117.00){\makebox(0,0)[cc]{0.08119...}}
\put(103.00,117.33){\makebox(0,0)[cc]{0.10769...}}
\put(133.00,117.00){\makebox(0,0)[cc]{0.12987...}}
\put(47.33,102.00){\makebox(0,0)[cc]{0.00614...}}
\put(74.33,102.00){\makebox(0,0)[cc]{0.05639...}}
\put(103.00,102.33){\makebox(0,0)[cc]{0.08088...}}
\put(133.00,102.33){\makebox(0,0)[cc]{0.10338...}}
\end{picture}
\newpage
\unitlength=1mm
\special{em:linewidth 0.4pt}
\linethickness{0.4pt}
\begin{picture}(149.67,146.67)
\put(7.00,73.67){\vector(1,0){142.67}}
\put(15.67,20.00){\vector(0,1){116.67}}
\put(72.33,73.67){\circle*{1.00}}
\put(21.67,131.67){\makebox(0,0)[cc]{$\phi(\Sigma)$}}
\put(144.67,77.00){\makebox(0,0)[cc]{$\Sigma$}}
\put(55.00,77.00){\makebox(0,0)[cc]{$\sigma_{k+1}$}}
\put(107.33,70.67){\makebox(0,0)[cc]{$\sigma_{k}$}}
\put(78.67,70.67){\makebox(0,0)[cc]{{\scriptsize $\Sigma_{k}(\mu,H)$}}}
\put(86.00,16.33){\makebox(0,0)[cc]{Fig. 5}}
\put(30.00,37.67){\circle*{0.33}}
\put(28.00,33.33){\circle*{0.33}}
\put(26.00,28.67){\circle*{0.33}}
\put(140.67,112.00){\circle*{0.33}}
\put(144.33,112.33){\circle*{0.33}}
\put(148.67,112.67){\circle*{0.33}}
\linethickness{1.00pt}
\bezier{368}(107.00,86.67)(100.00,107.67)(55.00,54.67)
\bezier{168}(107.00,86.67)(123.33,109.00)(137.67,111.67)
\bezier{208}(55.00,54.67)(50.67,70.67)(32.00,41.33)
\linethickness{0.2pt}
\put(55.00,73.67){\line(0,-1){1.33}}
\put(107.00,73.67){\line(0,1){1.33}}
\end{picture}
\newpage
\unitlength=1mm
\special{em:linewidth 0.4pt}
\linethickness{0.4pt}
\begin{picture}(147.67,145.33)
\put(6.00,76.33){\vector(1,0){141.67}}
\put(12.00,23.00){\vector(0,1){122.33}}
\put(76.67,24.33){\makebox(0,0)[cc]{Fig. 6}}
\put(19.00,135.33){\makebox(0,0)[cc]{$\phi(\Sigma)$}}
\put(144.00,81.33){\makebox(0,0)[cc]{$\Sigma$}}
\put(116.33,72.33){\makebox(0,0)[cc]{$\sigma_{k}$}}
\put(85.67,72.33){\makebox(0,0)[cc]{{\scriptsize $\Sigma_{k}(\mu,H)$}}}
\put(39.33,80.00){\makebox(0,0)[cc]{{\scriptsize $\Sigma_{k+1}(\mu,H)$}}}
\put(79.00,76.33){\circle*{1.00}}
\put(46.50,76.33){\circle*{1.00}}
\put(72.33,79.00){\makebox(0,0)[cc]{$\sigma_{k+1}$}}
\put(135.33,122.33){\circle*{0.33}}
\put(138.67,123.33){\circle*{0.33}}
\put(143.00,124.33){\circle*{0.33}}
\put(21.00,37.00){\circle*{0.33}}
\put(19.00,32.67){\circle*{0.33}}
\put(17.33,27.00){\circle*{0.33}}
\linethickness{1.00pt}
\bezier{92}(34.33,38.33)(28.33,48.67)(22.67,39.33)
\bezier{408}(116.33,105.67)(105.00,131.00)(73.00,64.33)
\bezier{96}(116.33,105.67)(123.67,118.33)(132.67,121.33)
\bezier{532}(73.00,64.33)(50.33,114.33)(34.33,38.33)
\linethickness{0.1pt}
\put(116.67,76.33){\line(0,1){1.00}}
\put(73.33,76.33){\line(0,-1){1.00}}
\end{picture}
\newpage
\unitlength=1mm
\special{em:linewidth 0.4pt}
\linethickness{0.4pt}
\begin{picture}(138.67,130.00)
\put(10.67,24.00){\vector(1,0){128.00}}
\put(18.67,12.67){\vector(0,1){117.33}}
\linethickness{0.1pt}
\put(60.33,24.00){\line(0,1){2.00}}
\put(72.67,24.00){\line(0,1){2}}
\put(91.00,24.00){\line(0,1){2.00}}
\put(11.00,123.67){\makebox(0,0)[cc]{$\Sigma(\mu,H)$}}
\put(10.67,94.33){\makebox(0,0)[cc]{$\scriptsize{\Sigma_0(H)}$}}
\put(60.67,19.00){\makebox(0,0)[cc]{$\mu_0$}}
\put(72.67,19.00){\makebox(0,0)[cc]{$\mu_1$}}
\put(91.00,19.00){\makebox(0,0)[cc]{$\mu_2$}}
\put(106.00,19.00){\makebox(0,0)[cc]{$\scriptsize{\tilde\mu_c(H)}$}}
\put(133.67,28.67){\makebox(0,0)[cc]{$\mu$}}
\put(80.33,8.00){\makebox(0,0)[cc]{Fig. 7}}
\linethickness{1pt}
\put(18.67,94.33){\line(1,0){42.00}}
\bezier{72}(60.33,85.67)(73.00,84.33)(73.00,79.00)
\bezier{128}(72.67,71.33)(90.33,66.00)(91.00,52.33)
\bezier{124}(91.00,42.67)(105.33,39.67)(104.00,24.00)
\end{picture}
\newpage
\unitlength=1.00mm
\special{em:linewidth 0.4pt}
\linethickness{0.4pt}
\begin{picture}(142.33,139.67)
\put(6.33,24.00){\vector(1,0){136.00}}
\put(15.33,12.67){\vector(0,1){127.00}}
\put(75.33,77.67){\circle*{0.67}}
\put(75.33,83.33){\circle*{0.67}}
\put(75.33,90.67){\circle*{0.67}}
\put(75.33,129.00){\circle*{0.67}}
\put(75.33,136.67){\circle*{0.67}}
\put(80.33,10.67){\makebox(0,0)[cc]{Fig. 8}}
\put(136.00,16.00){\makebox(0,0)[cc]{$H$}}
\put(9.67,50.00){\makebox(0,0)[cc]{$M$}}
\put(9.33,131.67){\makebox(0,0)[cc]{$\mu$}}
\put(75.33,36.00){\makebox(0,0)[cc]{$C_0$}}
\put(75.33,60.00){\makebox(0,0)[cc]{$C_1$}}
\put(75.33,111.67){\makebox(0,0)[cc]{$C_{n+1}$}}
\put(119.33,50.00){\makebox(0,0)[cc]{$l_0$}}
\put(119.33,77.00){\makebox(0,0)[cc]{$l_1$}}
\put(119.33,105.00){\makebox(0,0)[cc]{$l_n$}}
\put(123.33,132.33){\makebox(0,0)[cc]{$l_{n+1}$}}
\linethickness{1pt}
\bezier{300}(15.33,50.00)(36.67,73.33)(115.33,77.00)
\bezier{300}(15.33,50.00)(25.33,92.67)(115.33,104.67)
\bezier{400}(15.33,50.00)(20.33,120.33)(115.67,131.67)
\put(15.33,50.00){\line(1,0){100.33}}
\end{picture}
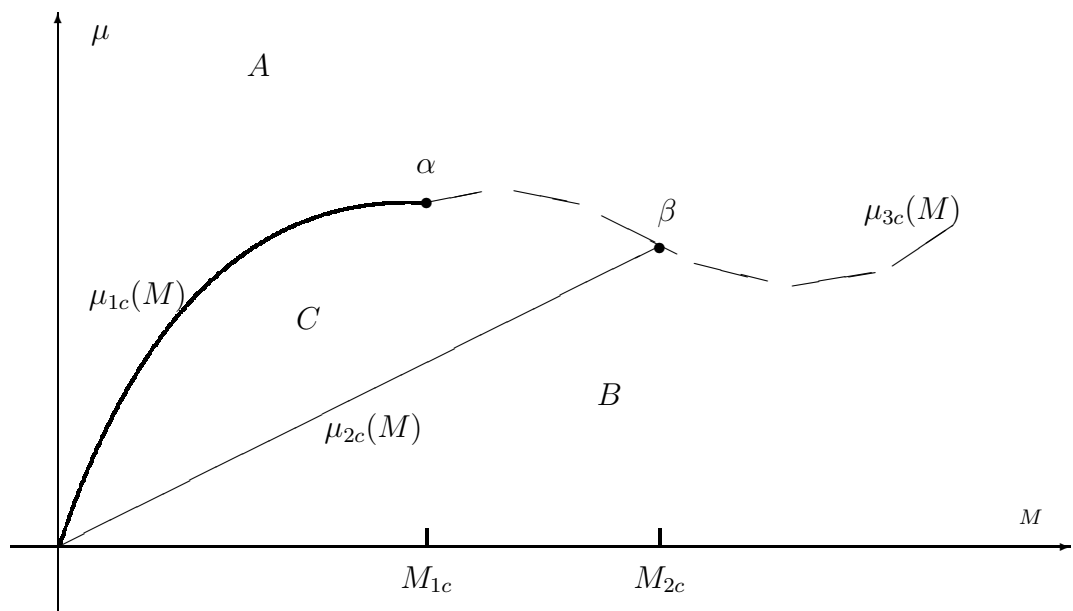
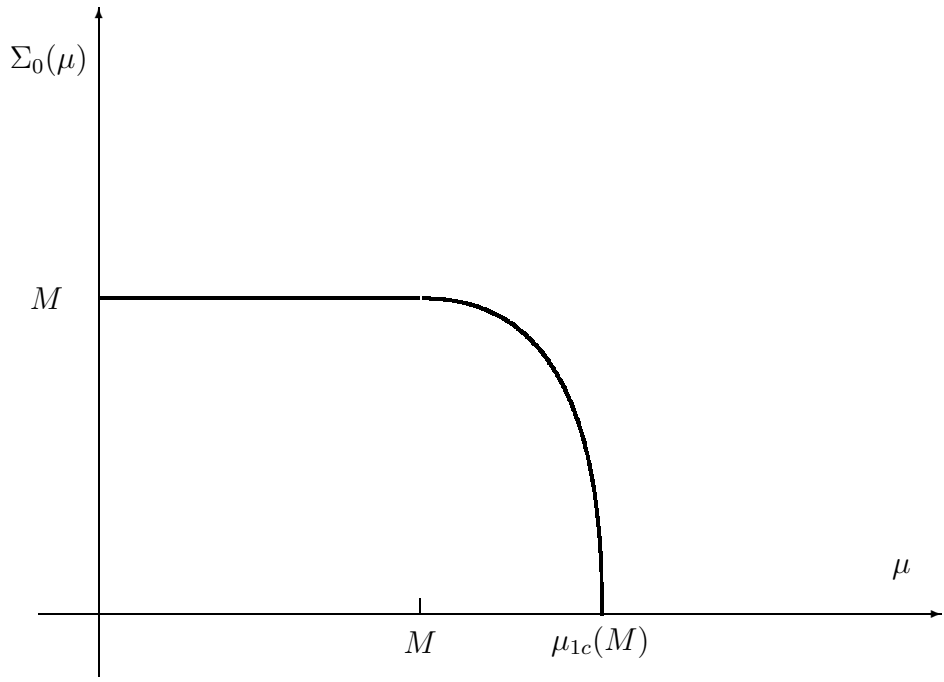
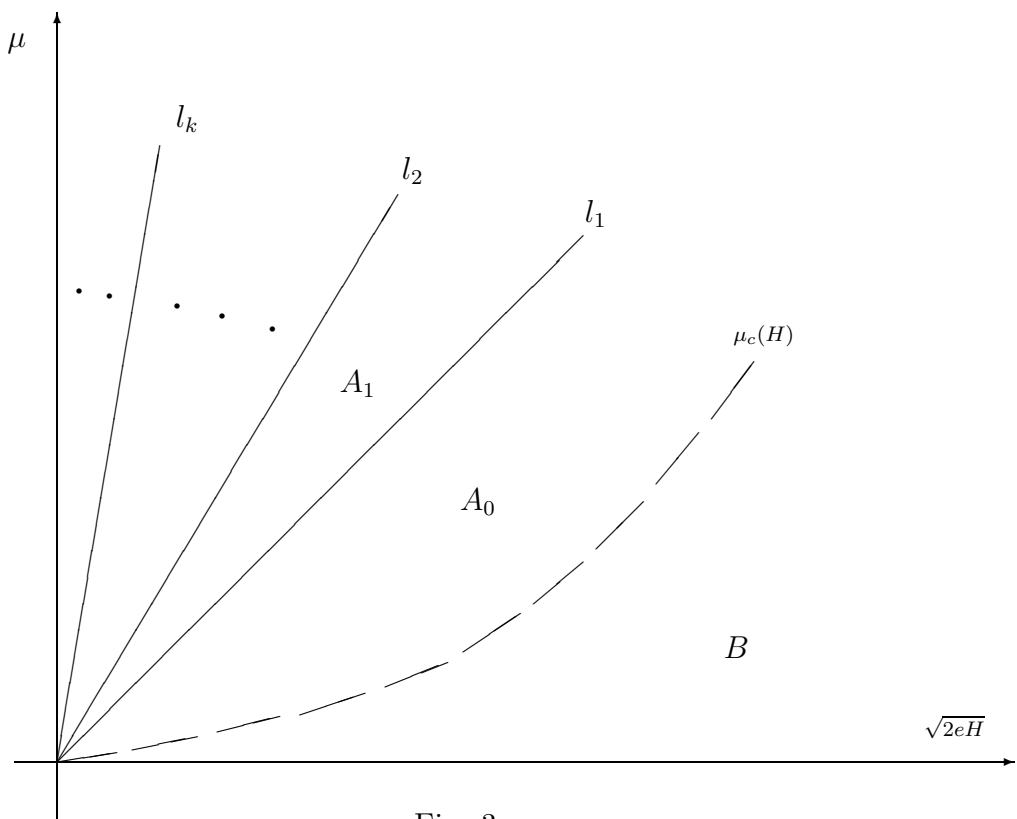
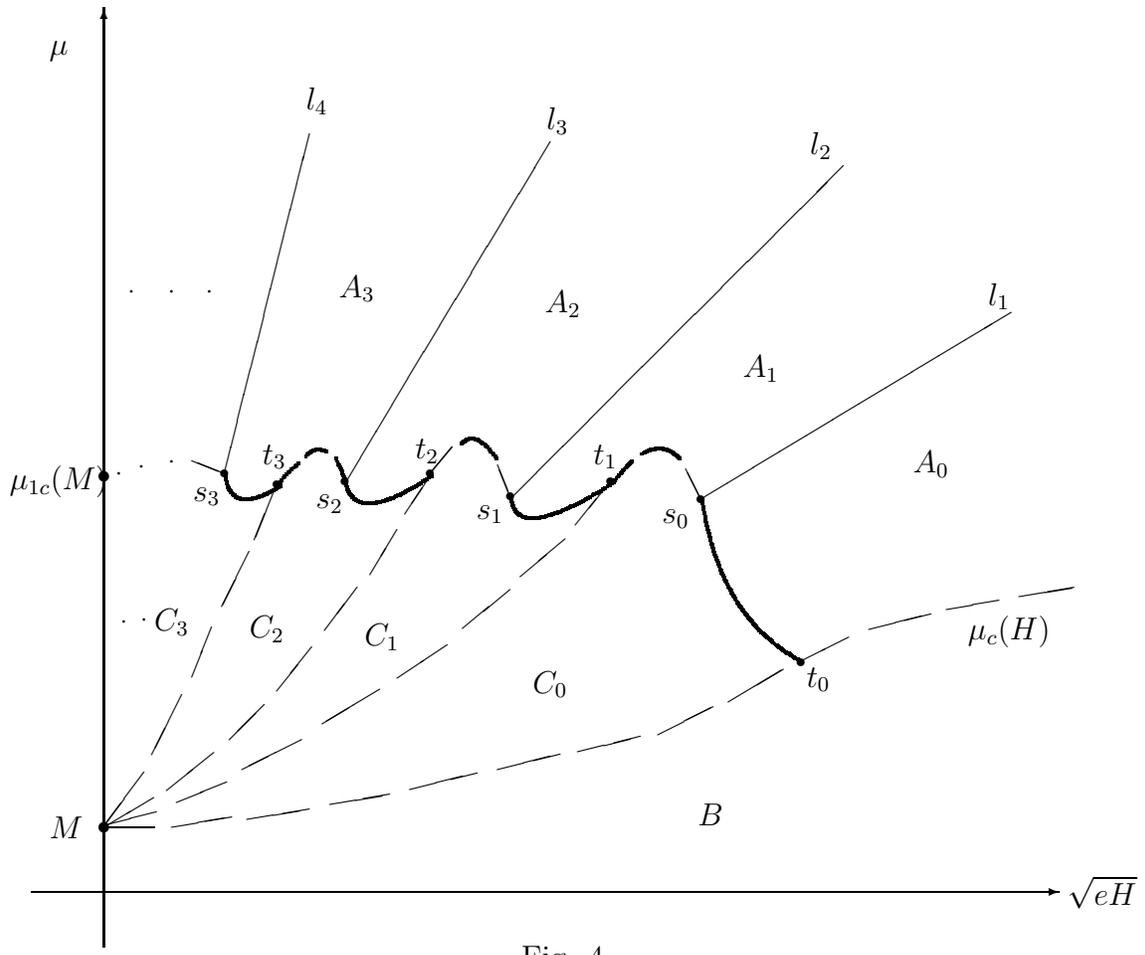
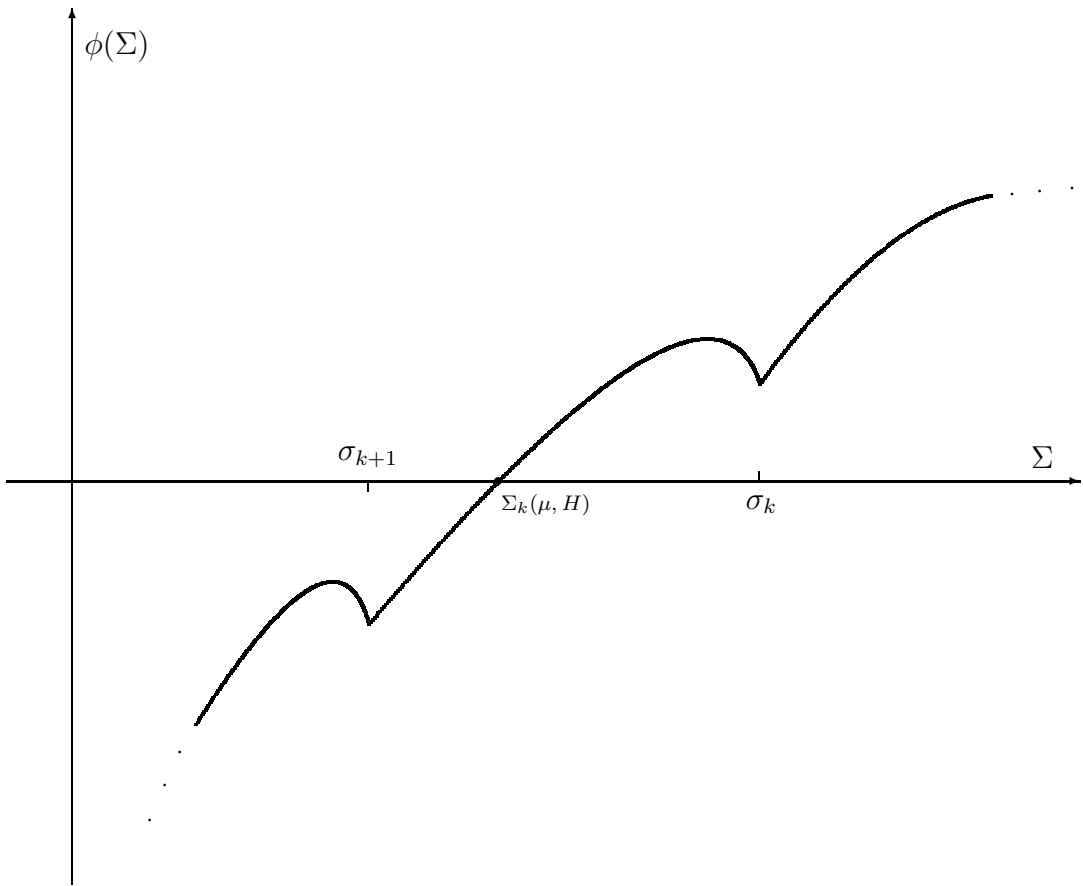
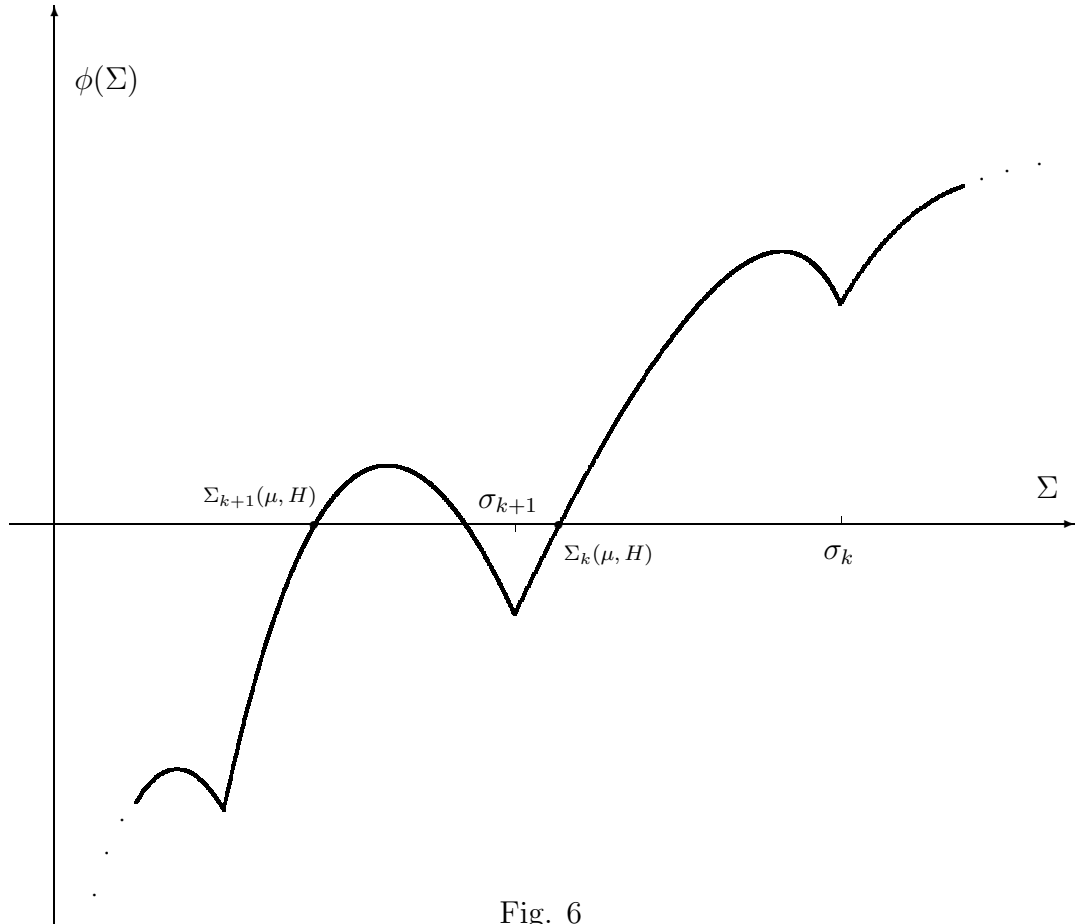
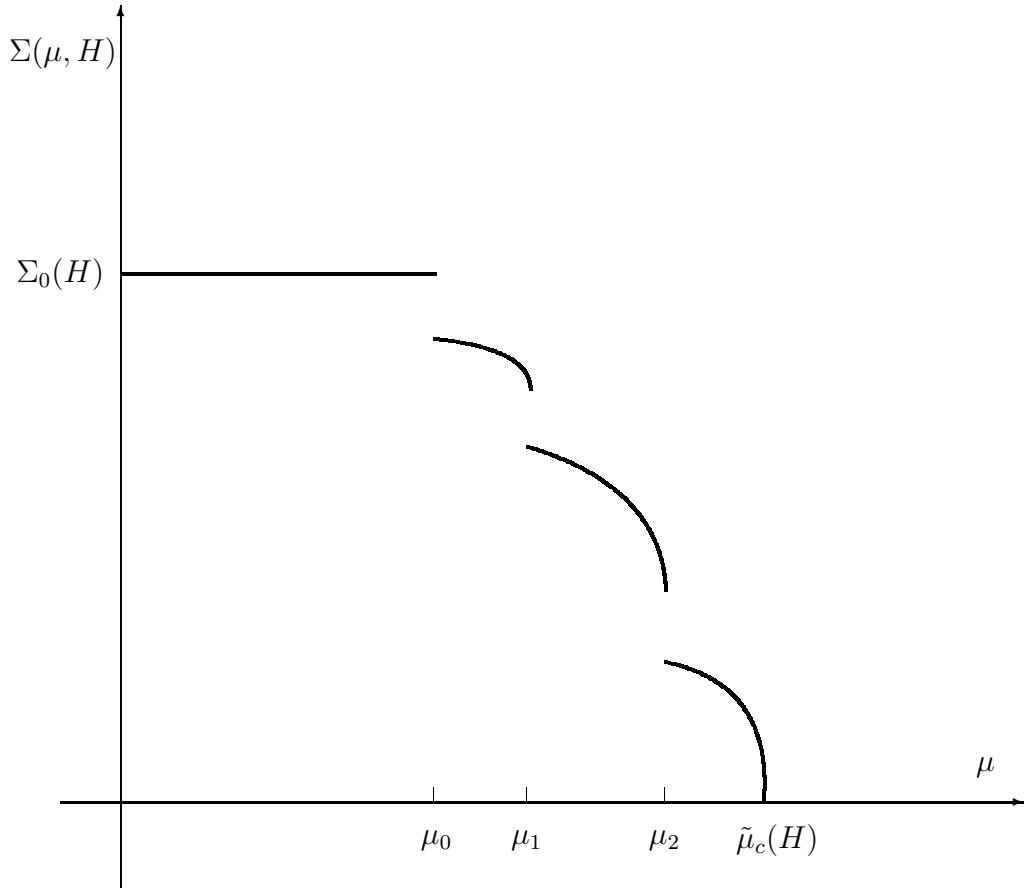
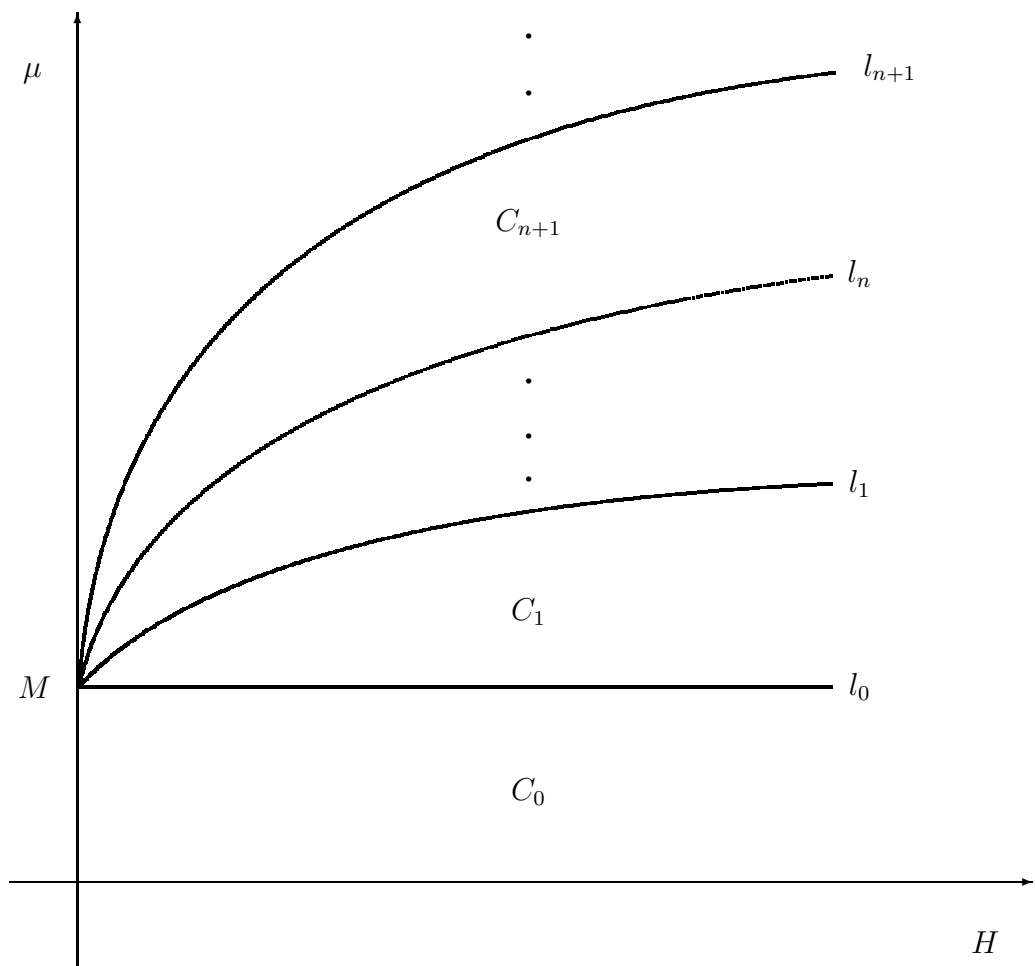
\end{document}